%% file: main.tex
\documentclass{aa}  

\usepackage{lscape}
\usepackage{graphicx}
\usepackage{array,multirow}
\usepackage{multirow}
\usepackage{txfonts}
\usepackage{gensymb}
\usepackage{natbib}
\usepackage{siunitx}
\bibpunct{(}{)}{;}{a}{}{,} 
\defcitealias{hilker2015}{Paper II}

\begin{document}

   \title{The Hydra I cluster core \\I. Stellar populations in the cD galaxy NGC 3311\thanks{Based on observations made with ESO Telescopes at the La Silla Paranal Observatory under programme ID 088.B-0448(B) PI Richtler.}}

   \author{C. E. Barbosa\inst{\ref{eso},\ref{usp}}\fnmsep\thanks{Corresponding author: {\tt carlos.barbosa@usp.br}} \and
              M. Arnaboldi\inst{\ref{eso},\ref{inaf}} \and
              L. Coccato\inst{\ref{eso}} \and
              M. Hilker\inst{\ref{eso}} \and
              C. Mendes de Oliveira\inst{\ref{usp}} \and
              T. Richtler\inst{\ref{concepcion}}
          }
\institute{
European Southern Observatory,  Karl-Schwarzschild-Stra\ss{}e 2, 85748, Garching, Germany\label{eso} 
\and
Universidade de S\~ao Paulo, IAG, Departamento de Astronomia, Rua do Mat\~ao 1226, S\~ao Paulo, SP, Brazil\label{usp}
\and
INAF, Osservatorio Astronomico di Torino, STrada Osservatorio 20, 10025 Pino Torinese, Italy\label{inaf}
\and
Departamento de Astronomia, Universidad de Concepci\'on, Concepci\'on, Chile\label{concepcion}      
}

\date{Received January 15, 2016; accepted.}

 
  \abstract
   {The history of the mass assembly of brightest cluster galaxies may be studied by the mapping the stellar populations at large radial distances from the galaxy centre, where the dynamical times are long and preserve the chemodynamical signatures of the accretion events.}   
   {To provide extended and robust measurements of the stellar population parameters in NGC~3311, the cD galaxy at the centre of the Hydra I cluster and out to three effective radius. We wish to characterize the processes that drove the build up of the stellar light at all these radii.}
   {We obtained the spectra from several regions in NGC~3311 covering an area of $\sim 3$ arcmin$^2$ in the wavelength range $4800\lesssim\lambda(\AA)\lesssim5800$, using the FORS2 spectrograph at the VLT in the MXU masking mode. We measured the equivalent width of seven absorption-features defined in the Lick/IDS system, that were modelled by single stellar populations to provide luminosity-weighted ages, metallicities and alpha element abundances.}
   {The trends in the Lick indices and the distribution of the stellar population parameters indicate that the stars of NGC~3311 may be divided in two radial regimes, one within and the another beyond one effective radius, $R_e = 8.4$ kpc, similar to the distinction between inner galaxy and external halo derived from the NGC~3311 velocity dispersion profile. The inner galaxy ($R\leq R_e$) is old (age $\sim 14$ Gyr), have negative metallicity gradients and positive alpha element gradients. The external halo is also very old, but have a negative age gradient. The metal and element abundances of the external halo have both a large scatter, indicating that stars from a variety of  satellites with different masses have been accreted. The region in the extended halo associated with the off-centred envelope at $0^\circ < \mbox{P.A.} < 90^\circ$ \citep{2012A&A...545A..37A} has higher metallicity with respect to the symmetric external halo.}
   {The different stellar populations in the inner galaxy and extended halo reflect the dominance of \textit{in situ} stars in the former and the accreted origin for the large majority of the stars in the latter. The low value of the velocity dispersion in the inner galaxy indicates that its stars are bound to the galaxy's gravitational potential, and the abundances and gradients suggest that the inner galaxy is formed in an outside-in scenario of merging gas-rich lumps, reminiscent of the first phase of galaxy formation. The external halo has a higher velocity dispersion, it is dynamically hotter than the galaxy and its stars are gravitationally driven by the cluster's gravitational potential. The stars in the external halo were removed from their parent galaxies, either disks with truncated star formation or the outer regions of early-type galaxies. Late mass accretion at large radii is now coming from tidal stripping of stars from dwarfs and S0 galaxies. These results provide supporting evidence to the recent theoretical models of formation of massive ellipticals as a two-phase process.}


   \maketitle
%
%
\newcolumntype{x}[1]{>{\centering\let\newline\\\arraybackslash\hspace{0pt}}p{#1}}
\section{Introduction} 

Brightest Cluster Galaxies (BCGs) are the giant early-type galaxies found at the core of galaxy clusters. BCGs often display extended, diffuse stellar halos, known as cD envelopes, which are believed to be composed of stars stripped from satellite galaxies interacting with the cluster's halo by tidal interactions and dynamical friction \citep{1972AJ.....77..288G,1979ApJ...231..659D}. Due to the long relaxation times  at large radii, stellar halos are not well-mixed and, therefore, preserve chemodynamical signatures of the past accretion events which have survived to present day and can be used to access the history of the galaxy mass assembly. 

The initial processes of galaxy formation through monolithic collapse do set  abundance gradients, which are modified by the subsequent history of accretion over time. Initial strong metallicity gradients arise naturally in galaxies formed by the collapse of gas clouds in the monolithic scenario \citep{1974MNRAS.166..585L} or by the merging of numerous gas rich sub-galaxies \citep{2004MNRAS.347..740K}. The subsequent evolution via major mergers is able to dilute metallicity gradients by the mixing of stars in the central regions of galaxies \citep{1980MNRAS.191P...1W,2004MNRAS.347..740K}, while minor mergers may form gradients at larger radii by the deposit of stars with different abundances \citep{1983MNRAS.204..219V,2015MNRAS.449..528H}. Observations in the last decade, such as the population of compact massive objects at high redshift \citep{2008ApJ...677L...5V,2014ApJ...788...28V} and the evolution of sizes and concentrations with redshift \citep{2009ApJ...699L.178N}, suggest that massive ellipticals are formed by different processes over time in the two-phase scenario \citep{2007MNRAS.375....2D,2010ApJ...725.2312O}. Galaxies are formed at high redshift ($z\gtrsim 3$) during star formation events triggered by rapid dissipative processes, such as cold accretion of gas through filaments \citep{2005MNRAS.363....2K,2009Natur.457..451D} or gas-rich major mergers \citep{2006ApJ...645..986R}, followed by an extended accretion of stars of smaller galaxies, usually at large galactic radii.

In the two-phase scenario, stars inhabiting a halo can be divided into two categories according to the sub-halo in which they have been formed: \textit{in situ} and accreted. The \textit{in situ} stars were formed in the most massive sub-halo of a galaxy during the first phase of galaxy formation, which resulted in stars with relatively high metallicity, due to their deep gravitational potential that can retain the metals ejected by supernovae \citep{2004ApJ...613..898T}, and high alpha element abundance due to the rapid time-scales of star formation \citep[see, e.g.,][]{2005ApJ...621..673T}. The accreted stars were formed in galaxies with a variety of masses and, consequently, have different abundances and ages. Spatially, \textit{in situ} stars are centrally concentrated and usually dominate the light in the central $\sim 5-10$ kpc of elliptical galaxies, contributing significantly to the stellar halo out to $\sim30$ kpc, while the accreted stars are less centrally concentrated and dominate the light in the outer regions \citep{2010ApJ...725.2312O,2013MNRAS.434.3348C}. From a dynamical point of view, \textit{in situ} stars have suffered violent relaxation processes in the early phases of the galaxy formation, while the accreted stars may be still unrelaxed, especially at the largest radii, and thus may exhibit an inhomogeneous spatial distribution which survived to the present day \citep{2015MNRAS.451.2703C}.

The observational test for this model of galaxy formation relies on the measurement of stellar population parameters in early-type galaxies out to large radii, but the low surface brightness in these regions makes spectroscopic measurements difficult. Nevertheless, current state-of-the art investigations seem in agreement with the predictions from the two-phase scenario. In the case of NGC~4889 in the Coma cluster, \citet{2010MNRAS.407L..26C} showed that there are different populations in the core and halo consistent with the idea of accreted stars dominating the light at radii $R > 18$ kpc. Similarly, differences between the inner and outer stellar populations have been found for other cD galaxies such as NGC~3311 \citep{2011A&A...533A.138C}, M49 \citep{2013ApJ...764L..20M}, M87 \citep[Virgo cluster,][]{2014MNRAS.439..990M} and NGC~6166 \citep[Abell 2199,][]{2015ApJ...807...56B}, and even in several non BCGs \citep{2014MNRAS.442.1003P}. However, these studies provide limited information about the distribution of the metal abundances and ages, because they were performed either by photometry, which has good spatial information but does not provide detailed abundances and ages, or by long-slit spectroscopy, which is limited to the slit position, but provides abundance and age information in detail. Ideally we would like to be able to collect the metal abundance and age information over the entire spatial extension of the galaxy light, for an extensive mapping of its physical properties.



In this work we study the galaxy NGC~3311, the cD of the Hydra I cluster, in order to provide the first bi-dimensional and large scale view of the stellar population of this system, which together with the detailed kinematic study (paper II) can shed further insights in a number of specific formation scenarios that have been recently proposed for this system. NGC~3311 has a radial velocity of $\sim$3800 km s$^{\rm -1}$ and a positive velocity dispersion profile \citep{2010A&A...520L...9V}, indicating that its extended halo is driven by the cluster potential at $R> 8$ kpc and may be composed of the debris shredded from satellite galaxies falling into the cluster's centre \citep{2011A&A...533A.138C}. Furthermore, such rising velocity dispersion profile seems to be asymmetric at large radii, as observed at different position angles \citep{2010A&A...520L...9V,2011A&A...531A.119R}. A scenario of on-going interactions and extended built-up of the stellar mass is also supported by the presence of multiple components in the line-of-sight velocity distribution of the planetary nebulae in NGC~3311 \citep{2011A&A...528A..24V}, of substructures and tails in the halo light distribution \citep{2012A&A...545A..37A} and by the distribution of dwarf galaxies around NGC 3311 \citep{2008A&A...486..697M}. 

In the accompanying paper on the kinematics of NGC~3311 \citep[\citetalias{hilker2015},][]{hilker2015}, we focus on the kinematic properties, discuss the implications for the mass profile of NGC~3311 and explore possible formation histories of the massive star clusters observed by \citet{2011A&A...531A...4M}. In the current paper, we present the spatial map of the stellar population parameters, ages and abundances, out to large radii. The article is organized as follow. In Section~\ref{sec:observations} we describe our data set and the methods of data reduction. In Section~\ref{sec:lick} we describe the measurements for the absorption features using Lick indices, and in Section~\ref{sec:ssps} we convert this information into physical parameters of the stellar populations. In Section \ref{sec:correspondence}, we discuss the correspondence of the stellar populations with kinematic and morphological structures, which are discussed in detail in Section \ref{sec:discussion}. In Section \ref{sec:implication} we discuss our main findings in line with recent observational and theoretical works and provide an updated view of the NGC~3311 halo formation based on recent works. Finally, we proceed to an overview and conclusion of this work in Section~\ref{sec:conclusion}. We assume a distance to the core of the Hydra cluster of 50.7 Mpc, calculated by the Hubble flow considering a radial velocity of 3777 km s$^{\rm -1}$ \citep{1999ApJS..125...35S} and $H_0=70.5$ km s$^{\rm -1}$ Mpc$^{\rm -1}$ \citep{2009ApJS..180..330K}. The adopted effective radius for NGC~3311 is $R_e= 8.4$ kpc, which is the mean value of the isophotal analysis in the V-band from \citet{2012A&A...545A..37A}.

\section{Observations and Data Reduction}
\label{sec:observations}
 
In this study we explore a new spectroscopic data set for the Hydra I cluster. The data was observed  at the Very Large Telescope (VLT) at Paranal, Chile, using the UT1 8.2m telescope with the FOcal Reducer/low dispersion Spectrograph 2 \citep[FORS2,][]{1998Msngr..94....1A} in the multi-object spectroscopic mode with the Mask Exchange Unit (MXU), obtained under ESO programme ID 088.B-0448B (PI: Richtler). 

\begin{figure}[!t]
\centering
\includegraphics[width=\linewidth]{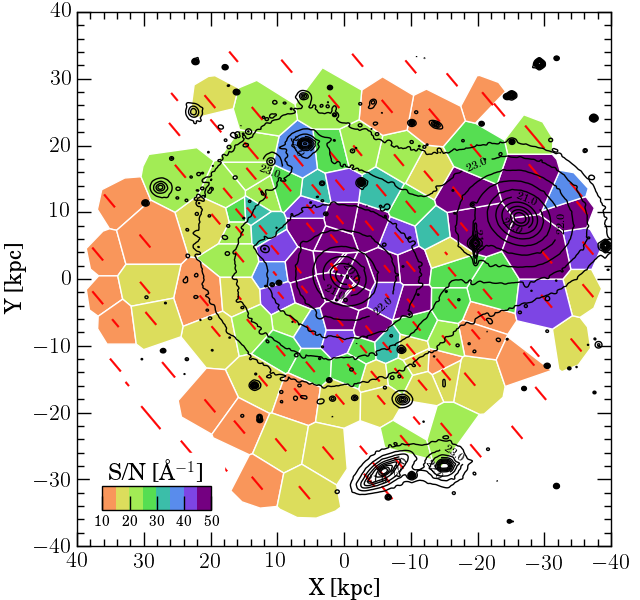}
\caption{Distribution of the signal-to-noise (S/N) of the scientific spectra in our sample. The position of the slits is shown as red bars, and the colours of the polygon, calculated by Voronoi Tessellation, indicate the S/N ratio according to the scale in the colour bar in the bottom left corner. Slits with white background were excluded from the analysis due to their low quality (S/N$<10$). Black lines show the V-band contours in the range from 20 to 23.5 mag arcsec$^{-2}$ in intervals of 0.5 mag arcsec$^{-2}$, from \citet{2012A&A...545A..37A}. {The geometry of the polygons is shown herefater in all the other maps in this study.}}
\label{fig:sn}
\end{figure}

NGC 3311 spans a large area of the sky, and thus its observation out to large galactocentric distances requires a field-of-view beyond the area of most integral field units available to date. In order to survey an area of 3$\times$3 arcmin$^{\rm 2}$ around NGC 3311, which translates into projected distances of $\sim 30$ kpc, we sampled regions of the stellar halo devoid of contamination of point sources with small slits, with typical size of 1"$\times$5", using 6 masks which sample the stellar halo in shells: cen1, cen2, inn1 inn2, out1 and out2. {We used the grism 1400V with the standard collimator to obtain a dispersion of $0.25 \AA$ pixel$^{-1}$, which gives a spectral resolution of $R=2100$ at $5200$\AA. The exposure times for the cen$_i$ and inn$_i$ masks covering the central region are $2\times1400$s, while it is $6\times1400$s for the out$_i$ masks. This observing strategy resulted in 135 slitlets dedicated to the observation of NGC~3311 and NGC 3309}. {The numbers of extracted spectra is smaller though, because we set a minimum signal-to-noise (S/N) of {10} in our observations, in order to constrain the stellar population parameters. A total of {117} spectra are used in our analysis. They are indicated by the red slits in Fig.~\ref{fig:sn}}. 

The reduction processes for extracting the scientific spectra, including bias subtraction, flat fielding correction, and wavelength calibration, were performed using custom  \textsc{IRAF} scripts. {We have used the long-slit spectrum of a spectrophotometric standard star, HD 102070, to test the effect of the flux calibration on the Lick indices, which are used for the analysis of the stellar populations. This test has shown that the median change in the equivalent width of the indices is of 0.5\% and, therefore, there is no need of flux calibrate our spectra.} The spectrum for sky subtraction of each slit was obtained simultaneously to the observations at slits in the periphery of the CCD, using the same x-axis position of the data, which ensured that science and sky spectra share the same wavelength range and exposure time for the subtraction. The wavelength range for spectra varies slightly as a function of the x-axis in the CCD, but in most cases the interval $4800\lesssim\lambda$ (\AA)$\lesssim5800$ is available. For consistency, we calculated the signal-to-noise ratio {per angstrom} (S/N) of each spectrum in the range $5200\leq\lambda$ (\AA) $\leq5500$, and the distribution of the S/N is presented by the polygons in Fig.~\ref{fig:sn}. In this case, and in all the maps shown in this work, we present the distribution of parameters using these polygons calculated using Voronoi tessellation and the V-band contours from observations in  \citet{2012A&A...545A..37A} in order to improve the visualization.

\section{Line strength indices}
\label{sec:lick}

\subsection{Equivalent widths of Lick indices}

The stellar population properties are studied by the analysis of spectral absorption features using line-strength indices in the Lick/IDS system \citep{1994ApJS...95..107W,1998ApJS..116....1T}, which consists
in calculating the equivalent widths (EW) of absorption features in a given central band amidst two pseudo-continuum side-bands. This process was carried out as follows.

We obtained the line-of-sight velocity distribution (LOSVD) for each spectrum using the \textsc{pPXF} code \citep{2004PASP..116..138C} considering four Gauss-Hermite LOSVD moments and additive polynomials of order 12 in order to compensate for the variations in the continuum. {The adopted stellar templates come from the Single Stellar Population (SSP) models of the MILES library \citep{2006MNRAS.371..703S}, with ages from 1 to 15 Gyr, and metallicities in the range $-0.7\leq$[Z/H]$\leq0.2$. They are computed for a Salpeter Initial Mass Function with slope of 1.3. The template spectra from the MILES library have a resolution of FWHM=$2.5$\AA, which is slightly larger than that our observations, FWHM=$2.1$\AA. We then convolved our spectra with a Gaussian filter to match the resolution of the MILES stellar library.} In most cases, the best fit was obtained by a linear combination of a small number of SSPs, but in a few cases a secondary nebular component was accounted for in the templates, which included emission lines for H$\beta$ (4861\AA), NI (5200\AA) and OIII (4957\AA\hspace{1pt} and 5007\AA\hspace{1pt} at fixed ratio of 1:3), that is subtracted off the observed spectrum, if present. {For the measurement of the Lick indices, we convolve our spectra to match the resolution of the Lick/IDS system, using Gaussian filters of varied resolution according to \citet{1997ApJS..111..377W}, before measuring the EWs using a custom Python code.} 

To correct for the intrinsic broadening of the absorption features, we used the ratio of the equivalent widths in the best fit from \textsc{pPXF} before and after the LOSVD convolution as in \citet{2010A&A...519A..95C}, using the relation

\begin{equation}
I = \frac{I'_0}{I'_{\rm LOSVD}}\cdot I_{\rm meas}
\label{eq:broadcorr}
\end{equation}

\noindent where $I'_{\rm LOSVD}$ and $I'_0$ are the indices measured in the best fit spectra from \textsc{pPXF} with and without the LOSVD convolution, and $I_{\rm meas}$ and $I$ are the measured and the corrected indices for the extracted spectra. 

\begin{figure*}[!t]
\centering
\includegraphics[width=0.495\linewidth]{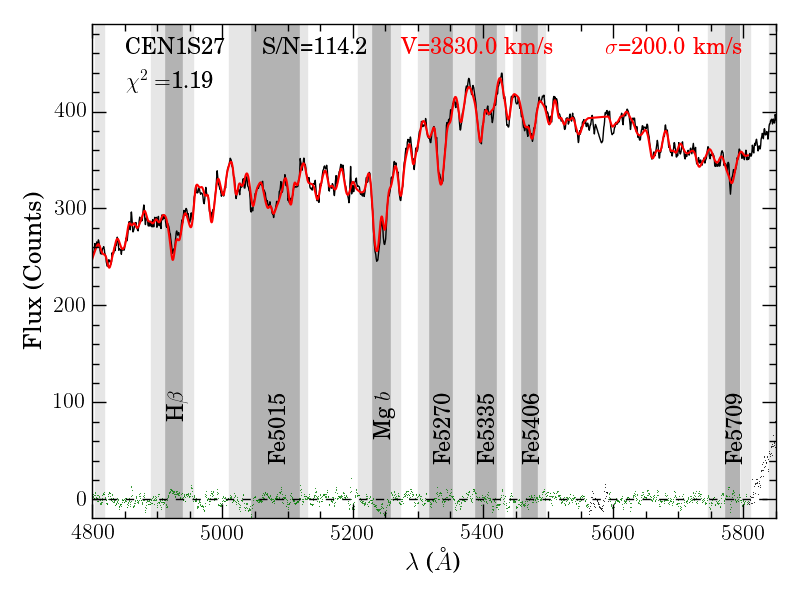}
\includegraphics[width=0.495\linewidth]{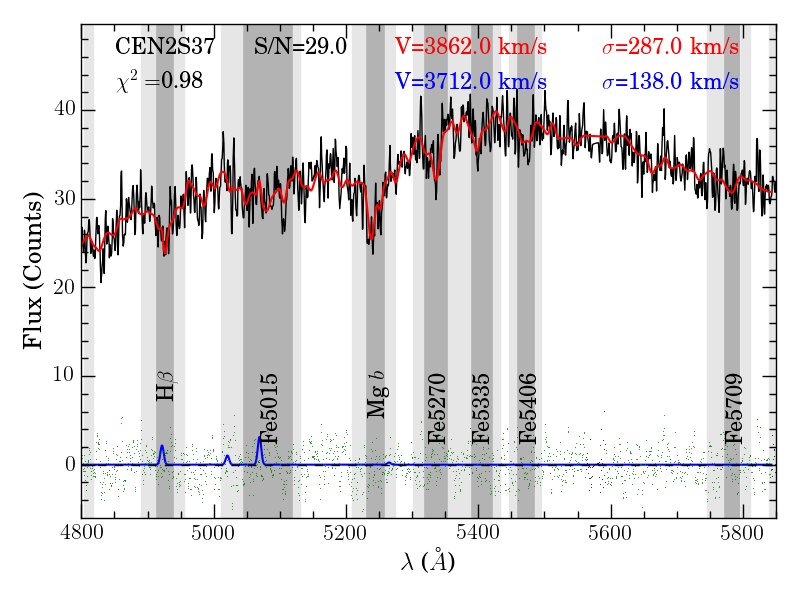}
\caption{Examples of the LOSVD fitting process with \textsc{pPXF} in our dataset. The sky subtracted spectra (black) are superposed by the best fit templates (red) obtained by a combination of SSP spectra from the MILES library \citep{2006MNRAS.371..703S}, emission lines and an additive polynomial of order 12, convolved with the line-of-sight velocity distribution with four moments. The best fit emission lines (blue) are subtracted from the spectra to avoid contamination on top of the absorption features. Residuals from the fits are shown in the bottom (green dots). The Lick indices are measured at the location of the vertical shades by the measurement of the equivalent width in central bands (dark grey) compared to the level of the two pseudo continuum side bands (light gray).}
\label{fig:fitexample}
\end{figure*}

Finally, we also corrected the indices for the instrumental offsets by the observation of the standard star HD~102070, for which Lick indices were determined by \citet{2007ApJS..171..146S}. Although not all indices are determined for HD~102070 in the spectral region of our data set, the relative small value of the correction in most cases and the good agreement with previous measurements for NGC~3311 from \citet{2012MNRAS.425..841L} assure us that any systematics are relatively small. Uncertainties for the Lick indices were estimated by Monte Carlo simulations with perturbations in the velocity dispersion and the addition of a bootstrapped noise, based on the residuals of the \textsc{pPXF} models. In Fig.~\ref{fig:fitexample}, we show typical examples of the extracted spectrum in our data set, including the best fit and residuals calculated with \textsc{pPXF} and the position of the Lick indices bands. In columns 5 to 11 of Table~\ref{tab:lick} we show the values for the corrected Lick indices of our extracted spectra.

\begin{longtab}
\begin{landscape}
\small
\begin{longtable}{cccccccccccccc}
\caption{\label{tab:lick} Equivalent width of Lick indices for the Hydra I cluster core and stellar population parameters.}\\
\hline\hline
ID & R & PA & S/N & H$\beta$ &  Fe5015 & Mg $b$ & Fe5270 & Fe5335 & Fe5406 & Fe5709 & $\log$ Age(yr) & [Z/H] & [$\alpha$/Fe]\\
   & (kpc) & ($^{o}$) &  & (\AA) &  (\AA) & (\AA) & (\AA) & (\AA) & (\AA) & (\AA) & (dex) & (dex) & (dex)\\
(1) & (2) & (3) & (4) & (5) & (6) & (7) & (8) & (9) & (10) & (11) & (12) & (13) & (14) \\
\hline
\endfirsthead
\caption{continued.}\\
\hline\hline
ID & D & PA & S/N & H$\beta$ &  Fe5015 & Mg $b$ & Fe5270 & Fe5335 & Fe5406 & Fe5709 & $\log$ Age(yr) & [Z/H] & [$\alpha$/Fe]\\
   & (kpc) & ($^o$) &  & (\AA) &  (\AA) & (\AA) & (\AA) & (\AA) & (\AA) & (\AA)& (dex) & (dex) & (dex)\\
(1) & (2) & (3) & (4) & (5) & (6) & (7) & (8) & (9) & (10) & (11) & (12) & (13) & (14)\\
\hline
\endhead
\hline
\endfoot
\input{lick.tex}
\end{longtable}
\tablefoot{(1) Identification of the spectrum. (2) Projected distance of the centre of the slit to the centre of NGC 3311 in kpc, assuming the distance to the Hydra I cluster is 50 Mpc. (3) Position angle of the centre of the slit to the centre of NGC 3311. North is 0$^{o}$, East is 90$^{o}$. (4) Signal to noise ratio. (5-11) Equivalent widths of Lick indices and their respective uncertainties. (12-14) Single stellar population parameters calculated from the Lick indices using models of \citet{2011MNRAS.412.2183T}.}
\end{landscape}
\end{longtab}

\subsection{Effect of systematic errors due to sky subtraction}
\label{sec:skysys}

Elliptical galaxies usually present large variances in their stellar population properties at large radii, such as observed in high S/N integral field observations from the SAURON survey \citep{2006MNRAS.369..497K,2010MNRAS.408...97K}, and our results show similar trends. {One question is whether varying S/N from the inner regions to large radii may be the primary source of scatter in the derived stellar population parameters, as function of radius.} In order to verify whether this result is related to different S/N, we performed the following test. Since the main source of uncertainty is the sky subtraction, we performed several measurements of the Lick indices with different values of under/over sky subtraction at the level of $\pm 1$\% of the total sky in each spectrum. This percentage was obtained by a visual inspection of our spectra after the inclusion of this systematic error, which showed that errors of this magnitude would leave easily recognizable features resembling the sky spectra in the 2D spectra of low signal-to-noise spectra (S/N$\sim 15$). 

{The result of this test is presented in Fig.~\ref{fig:lick_mad}, where the difference of the EW of the indices, $\delta I_{Lick}$, is plotted as a function of radius. The deviations are calculated using a running standard deviation, indicated by the grey shaded area. For a typical spectrum with S/N$\sim20$, 
the variations of the measured EWs for the Lick indices are $\delta$H$\beta$ =0.21 \AA (12\%), $\delta$Fe5015 =0.43 \AA (25\%), $\delta$Mg $b$ =0.35 \AA (11\%), $\delta$Fe5270 =0.22 \AA (18\%), $\delta$Fe5335 =0.30 \AA (8\%), $\delta$Fe5406 =0.20 \AA (10\%) and $\delta$Fe5709 =0.06 \AA (6\%). The same test is also performed for the stellar population parameters, using the methods described in Section~\ref{sec:ssps}, resulting in changes of $\delta$log Age (years)=0.05 dex, $\delta$[Z/H]=0.48 dex, $\delta$[$\alpha$/Fe]=0.14 dex and $\delta$[Fe/H]=0.45 dex (see Fig.~\ref{fig:ssps_std}). In the following analysis, we include the results of this test in Figs.~\ref{fig:radindices} and \ref{fig:ssp_gradients}, when we compare the dispersion due to systematic effects caused by the sky subtraction with the scatter of the measured Lick indices and stellar population parameters.}

\subsection{Spatial distribution and gradients of the Lick indices}

\begin{figure*}[!htb]
\centering
\includegraphics[width=\linewidth]{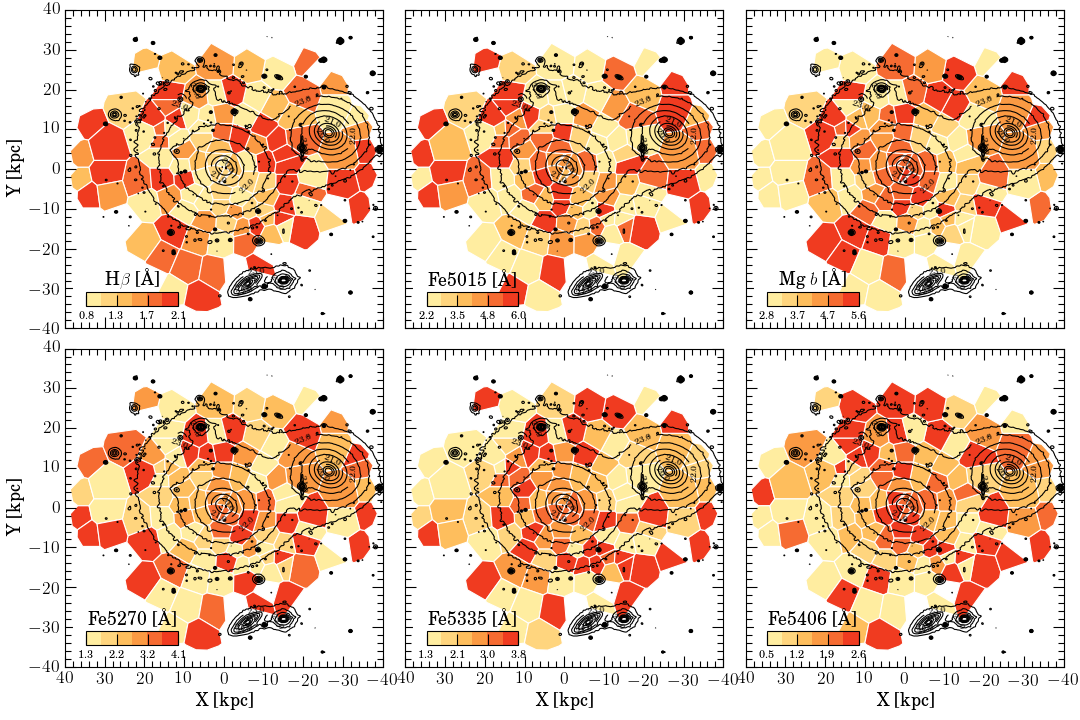}
\caption{Spatial distribution of the equivalent width of six Lick indices: H$\beta$, Fe5015, Mg $b$, Fe5270, Fe5335 and Fe5406. Black lines indicate the contours of the $V$-band image from \citet{2012A&A...545A..37A} between 20 to 23.5 mag arcsec$^{\rm -2}$ in steps of 0.5 mag arcsec$^{\rm -2}$. { These maps illustrate the presence of an homogeneous inner region, and a large scatter at large radii, for all indices.}}
\label{fig:mapindices}
\end{figure*}

The maps of the values for six measured Lick indices are presented in Fig.~\ref{fig:mapindices}, including H$\beta$, Fe5015, Mg $b$, Fe5270, Fe5335 and Fe5406. The index Fe5709 is not displayed in the figure because it is measured only in a fraction of our spectra, due to issues with the red continuum for some observational masks, but it is also used  for the modelling of stellar populations whenever possible. These maps indicate that each index has a particular pattern, and the central region of NGC~3311 stands out by having a relatively weaker H$\beta$ absorption and stronger metal line indices than the outskirts.  

\begin{figure}[!htb]
\centering
\includegraphics[width=0.95\linewidth]{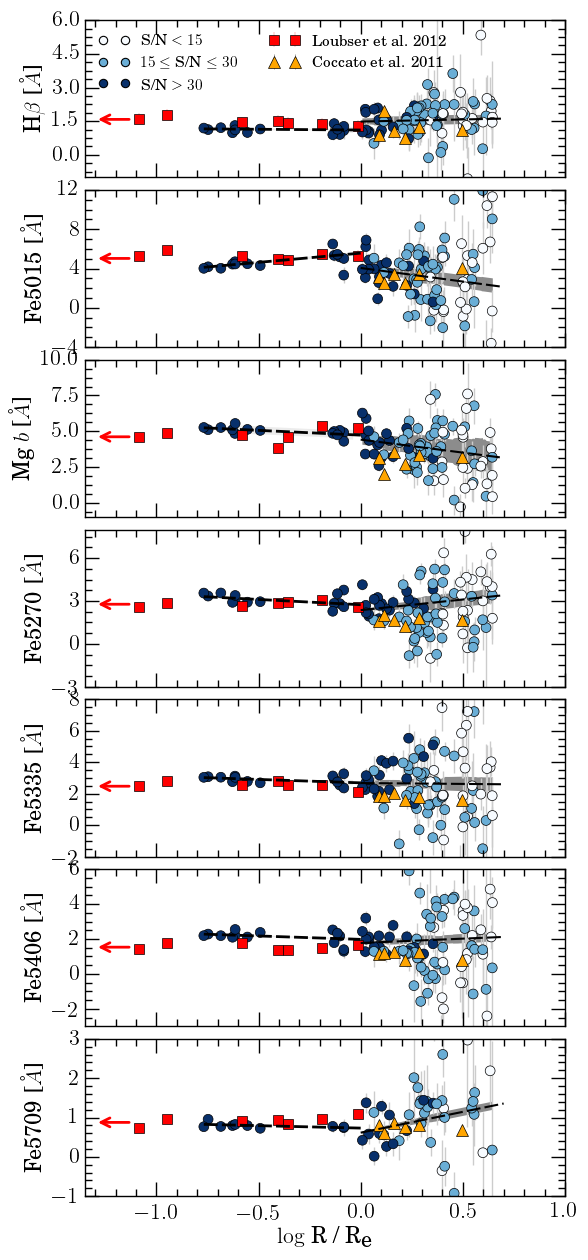}
\caption{Lick indices as a function of the distance from the centre of NGC~3311. {The EWs measured for our spectra are shown by full circles, with different colors according to their S/N, as indicated in the top panel.} Long-slit data from the literature is also displayed, including \citet{2011A&A...533A.138C} as orange triangles, \citet{2012MNRAS.425..841L} as red squares and  \citet{2009MNRAS.398..133L} as red arrows. The two black dashed lines indicate the gradients {obtained by linear regression in the two regions corresponding to the inner galaxy and the external halo. The grey shaded areas around the dashed lines represent the variance in the equivalent width generated by a systematic error in the sky subtraction by $\pm 1$\%. The different gradients indicate different mechanisms for the assembly of the stellar halos.}}
\label{fig:radindices}
\end{figure}

The large scale behavior of the Lick indices can be observed in Fig.~\ref{fig:radindices}, where the Lick indices are displayed as a function of the projected galactic log-radial distance from the centre of NGC~3311, in units of effective radius. The results from our measurements are shown as circles of different colours according to the signal-to-noise of the spectra. We also include a comparison with the long-slit data from the literature including \citet{2009MNRAS.398..133L}, \citet{2011A&A...533A.138C}, and \citet{2012MNRAS.425..841L}. Our indices are not only in good agreement with the literature, but also complement the data from \citet{2012MNRAS.425..841L} by extending the radial coverage. 

One important characteristic of Fig.~\ref{fig:radindices} is the large scatter in the external halo, as we already commented in Section~\ref{sec:skysys}. This scatter is larger than that caused by the systematic effects of over/under subtraction of the sky by $\pm1$\%, shown in the figure by the grey shaded areas. Hence one also expects a large true scatter in the abundance and age distributions at large radii. Further evidence for the intrinsic nature of the scatter at $R\sim R_e$ and outwards is the significant variance of the stellar population parameters derived for the highest S/N spectra in the data set, shown by the dark blue circles. 

Another important feature of Fig. \ref{fig:radindices} is the presence of a break in the radial gradients, from the inner to the outer profiles. This break occurs at about one effective radius ($R_e=8.4$ kpc), which is close to the expected transition between the \textit{in situ} and accreted components according to models \citep[e.g.][]{2015MNRAS.451.2703C}. Therefore, to quantify the mean trends of the Lick indices we divided them into two radial regions, the inner galaxy ($R\leq R_e$) and the extended halo ($R > R_e$). In each of these regions, we measured the radial gradients using the equation

\begin{equation}
I_{\rm{Lick}}(X) = I_{\rm{Lick}}(0) + \Delta I_{\rm{Lick}} \cdot X \mbox{,}
\label{eq:gradients}
\end{equation}

\noindent where $X=\log ({R/R_e})$ is the logarithm of the projected galactocentric distance $R$, normalized by the effective radius $R_e$, $I_{\rm{Lick}}(X)$ are the corrected Lick indices as a function of $X$, and $\Delta I_{\rm{Lick}}$ is the calculated gradient. { The gradient is determined from a $\chi^2$ minimization weighted by the uncertainties of the data points. We do not remove any outliers, but we implement a geometric selection by excluding the spectra that are i) very close to NGC~3309 and ii) on top of HCC~007. The coefficients of the linear regressions are presented in Table~\ref{tab:lickgrad} and the best fit is shown by the black dashed lines in Fig.~\ref{fig:radindices}.}

\begin{table}
\caption{{Linear regression coefficients for the Lick indices as function of radius}}             
\label{tab:lickgrad}   
\centering                         
\begin{tabular}{cx{1.5cm}x{1.5cm}x{1.5cm}x{1.5cm}}        
\hline\hline                 
 &  \multicolumn{2}{x{3cm}}{Inner galaxy (1)} & \multicolumn{2}{x{3cm}}{Extended halo (2)}\\
       & $I_{\rm{Lick}}(0)$ & $\Delta I_{\rm{Lick}}$ & $I_{\rm{Lick}}(0)$ & $\Delta I_{\rm{Lick}}$ \\
Index & (\AA)  & (\AA / dex)& (\AA) & (\AA / dex)\\ 
\hline                        
H$\beta$  & 1.1$\pm$0.1 & -0.1$\pm$0.2 & 1.5$\pm$0.1 & 0.2$\pm$0.5\\
Fe5015  & 5.6$\pm$0.3 & 1.9$\pm$0.5 & 4.1$\pm$0.4 & -2.8$\pm$1.3\\
Mg $b$  & 4.7$\pm$0.1 & -0.7$\pm$0.2 & 4.4$\pm$0.2 & -1.8$\pm$0.8\\
Fe5270  & 2.7$\pm$0.2 & -0.7$\pm$0.4 & 2.4$\pm$0.2 & 1.5$\pm$0.9\\
Fe5335  & 2.7$\pm$0.1 & -0.4$\pm$0.2 & 2.7$\pm$0.2 & -0.1$\pm$0.9\\
Fe5406  & 2.0$\pm$0.2 & -0.4$\pm$0.3 & 1.8$\pm$0.2 & 0.5$\pm$0.8\\
Fe5709  & 0.7$\pm$0.1 & -0.1$\pm$0.1 & 0.6$\pm$0.1 & 1.1$\pm$0.6\\
\hline  \hline                                 
\end{tabular}
\tablefoot{ (1) Linear regression coefficients for the Lick indices as function of radius in the inner galaxy ($R\leq R_e$), derived according to equation \eqref{eq:gradients}, where $I_{\rm{Lick}}(0)$ indicates the value of the indices at one effective radius and $\Delta I_{\rm{Lick}}$ indicates the gradients. (2) Same as (1) for the external halo ($R> R_e$) region.}
\end{table}

The main result of this section is summarized in Fig.~\ref{fig:radindices}. The distribution of the measured values for the Lick indices in the inner galaxy are different from the distribution of the values of the same indices in the external halo. The break occurs at $R = R_e$ for all indices. This implies that independent of the stellar population models that may translate the equivalent widths of the absorption features into other physically relevant quantities, there is going to be a break in the stellar population properties with different values for the inner galaxy and the external halo.

\section{Stellar populations}
\label{sec:ssps}

We adopted the model from \citet{2011MNRAS.412.2183T} to derive luminosity-weighted stellar population parameters -- age, total metallicity ([Z/H]) and alpha element abundance ([$\alpha$/Fe]) --, which are computed using a Salpeter initial stellar mass function. {} Also, we estimated the iron abundance using the relation \citep{2003MNRAS.339..897T}

\begin{equation}
\mbox{[Z/H]} = \mbox{[Fe/H]} + A \mbox{[$\alpha$/Fe]}
\label{eq:abundances}
\end{equation}

\noindent considering the factor $A=0.94$ \citep{2000AJ....120..165T}. All those parameters will help to understand the mass assembly history of stars in NGC 3311. The inferred ages may constrain the time since the last burst of star formation, the metallicities are a proxy for the mass of the parent halo where such stars were formed and the alpha element abundances set constraints on the star formation time-scales \citep{2005ApJ...621..673T}. 

Before we describe in detail the methods to compute the stellar population parameters, we list here distinct features of this modelling approach. We assume that the single stellar population (SSP) approximation is valid locally for each of our spectra, while we may be observing composite stellar populations (CSP). Each of the SSP parameters is sensitive to a different stellar population, which may or may not be the one dominating the stellar mass. In our discussion, we are going to use the interpretation of the stellar population parameters according to \citet{2007MNRAS.374..769S}, which relate SSP-equivalent parameters to CSPs as follows. Ages are biased towards the age of the youngest stellar populations, even in cases where they have only a small fraction of the mass, and therefore spectra with very low ages are not considered to be young, but instead old populations with a contribution of young stars. In case of the element abundances, these reflect the distribution of the most massive components in the composite stellar populations and, therefore, approximately reflect the mass-weighted properties.    

\subsection{Determination of SSP properties using Markov Chains}

We used Monte Carlo Markov Chains \citep[MCMC, ][]{Markov1913} to obtain SSP parameters, $\theta$=($\log $Age, [Fe/H], [$\alpha$/Fe]) from the information contained in the set of Lick indices, $D$, available for each spectrum. A complete discussion of the MCMC method can be obtained elsewhere \citep[e.g.,][]{mackay2003,2012psa..book.....W}, and here we only summarize the main concepts. The idea is to infer the posterior probability distribution of the parameters given the data, $p(\theta | D)$, using Bayes' theorem, 

\begin{equation}
p(\theta | D) \propto p(\theta) p(D | \theta)\mbox{,}
\label{eq:bayes}
\end{equation}

\noindent where $p(\theta)$ is the prior distribution and $p(D | \theta)$ is the likelihood distribution. We assume that the priors are uniform within the ranges of the models from \citet{2011MNRAS.412.2183T} extrapolated in metallicity, $0.1\leq$ Age(Gyr)$\leq15$, $-2.25\leq$ [Z/H] $\leq 0.90$ and $-0.3\leq$ [$\alpha$/Fe] $\leq 0.5$, and that the likelihood distributions are Gaussian functions with standard deviations equal to the uncertainties of the measurements of the Lick indices. The calculations of the samples were performed with the \textsc{PyMC} package \citep{pymc} using the Metropolis-Hastings algorithm with SSP models linearly interpolated for sub grid resolution. 

\begin{figure*}[!htb]
\centering
\includegraphics[width=0.495\linewidth]{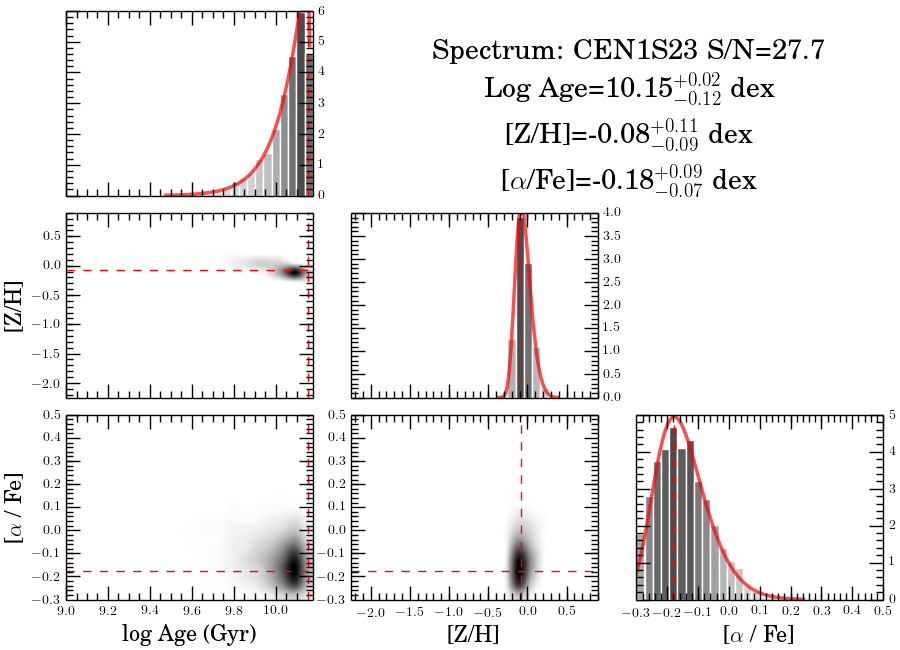}
\includegraphics[width=0.495\linewidth]{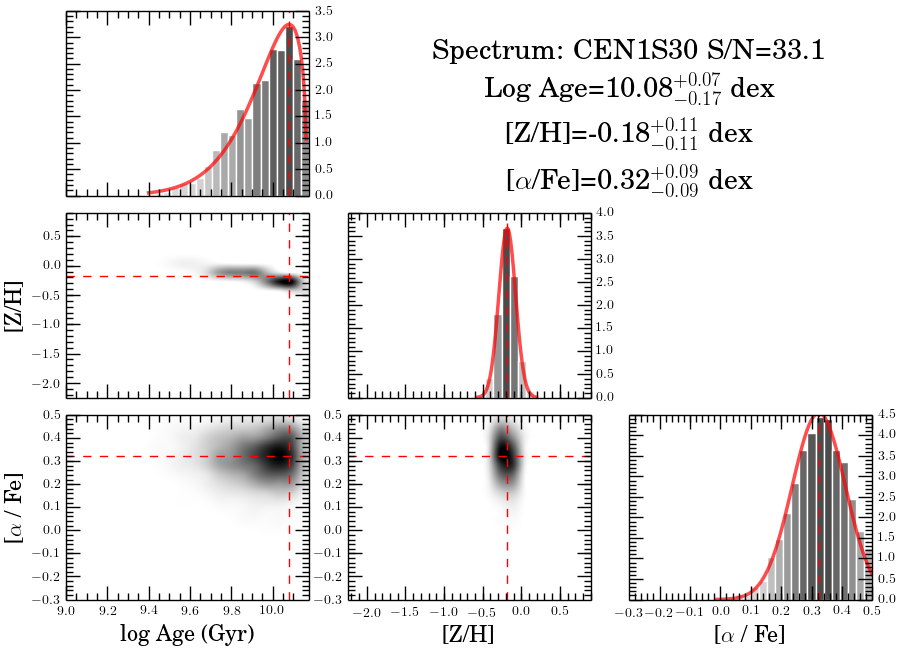}
\caption{Examples of the posterior distributions for two spectra using the MCMC method. The panels in the main diagonal show histograms of the marginalized distributions, which are used for the determination of the representative values and their uncertainties. The thick red lines show the best fit to the posterior samples, either a Gaussian or a Generalized Extreme Value function. Projections of the posterior distributions in the lower left panels show the correlations among parameters. {The histograms in the upper right corner of each panel indicate the maximum probabilities and the 1$\sigma$ deviations of the stellar population parameters in each case.}}
\label{fig:mcmcexample}
\end{figure*}

Fig.~\ref{fig:mcmcexample} illustrates the posterior distribution in two cases. The panels in the lower left show projections of the parameter space which indicate that parameters are not independent from each other. In particular, a small age-metallicity degeneracy is present \citep{1994ApJS...95..107W}. However, the relevant statistics are obtained by the marginalization of the parameters, indicated by the histograms, which already take these effects into account. For symmetric posterior distributions, such as the metallicities in both examples of Fig.~\ref{fig:mcmcexample}, we fit a simple Gaussian in order to obtain the mode and the standard deviation. However, in several cases the distributions are skewed towards the limits of the models, such as the ages in Fig.~\ref{fig:mcmcexample}, which required a fit of a Generalized Extreme Value (GEV) function. In these cases, we have used the mode as relevant statistic, and the uncertainties are calculated by determining two iso-probability values encompassing the maximum probabilities for which the posteriori distribution integrals add up to 68\% of the area under the curve, similarly to the $1\sigma$ deviations of a Gaussian distribution. 

The results of this analysis are presented in columns 12 to 14 of Table~\ref{tab:lick}. In low S/N regimes, the posterior distributions become flat due to the larger uncertainties in the Lick indices. To avoid such unconstrained parameters in the analysis, we have set a maximum limit of 0.2, 0.7, 0.2 and 0.8 dex in the uncertainties of $\log$ (Age), [$\alpha$/Fe], [Z/H] and [Fe/H] respectively.  

\subsection{Spatial distribution and gradients of SSP parameters}

\begin{figure*}
\centering
\includegraphics[width=\linewidth]{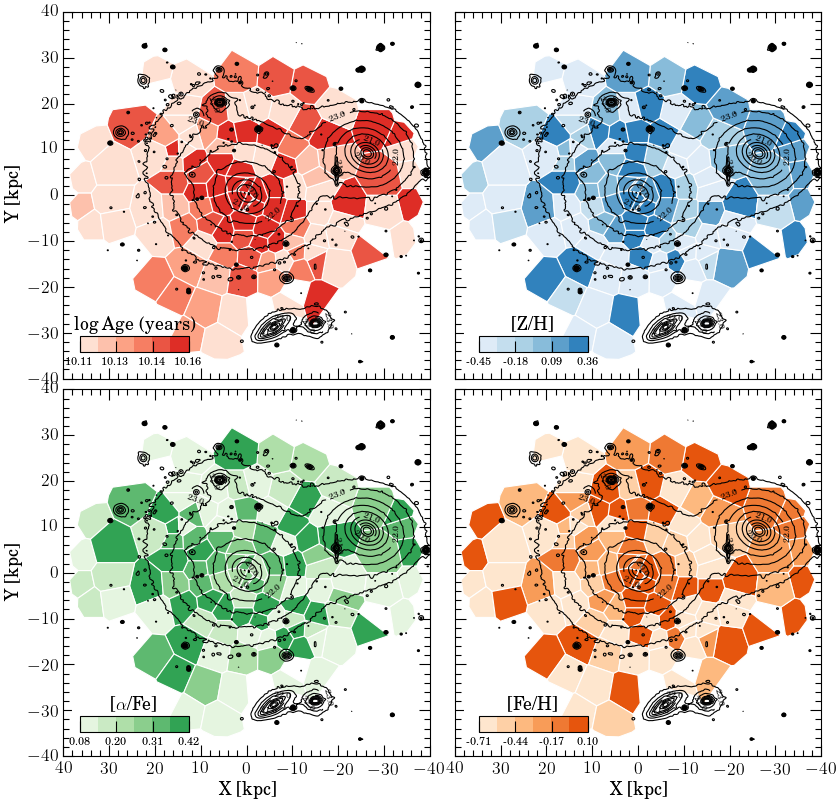}
\caption{Maps of the modeled luminosity-weighted stellar population properties: ages (top left), total metallicity (top right), alpha-element abundance (bottom left) and iron abundance (bottom right). Black lines display the $V$-band surface brightness contours from \cite{2012A&A...545A..37A} in the range from 20 to 23.5 in steps of 0.5 mag arcsec$^{-2}$. Colours indicate the SSP equivalent parameters according to the colour-bars in the bottom left section of each panel. {The inner galaxy ($R< \sim$10 kpc) is characterized by homogenous old age, high metallicity and super solar [$\alpha/$Fe]; the outer halo shows a more complex behavior; see extended discussion in Section~\ref{sec:discussion}.}}
\label{fig:ssp_lick_models}
\end{figure*}

In Fig.~\ref{fig:ssp_lick_models}, we present the spatial distribution of the four stellar population parameters, i.e. $\log (\mbox{Age})$, [Z/H], [$\alpha$/Fe] and [Fe/H]. As observed in the distribution of Lick indices, the large scale distribution of the stellar populations is scattered, predominantly in the outer regions, while the inner galaxy displays a more regular behaviour. The central region shows rather old ages and super solar abundances. The external halo is much more diverse, presenting some large scale structures uncorrelated to the isophotes. However, considering the low spatial resolution of our observations, the relative small S/N in several of these spectra, and the expected spread in stellar populations, we consider that most of the apparent structures may be explained by random fluctuations.  

\begin{figure*}
\centering
\includegraphics[width=\linewidth]{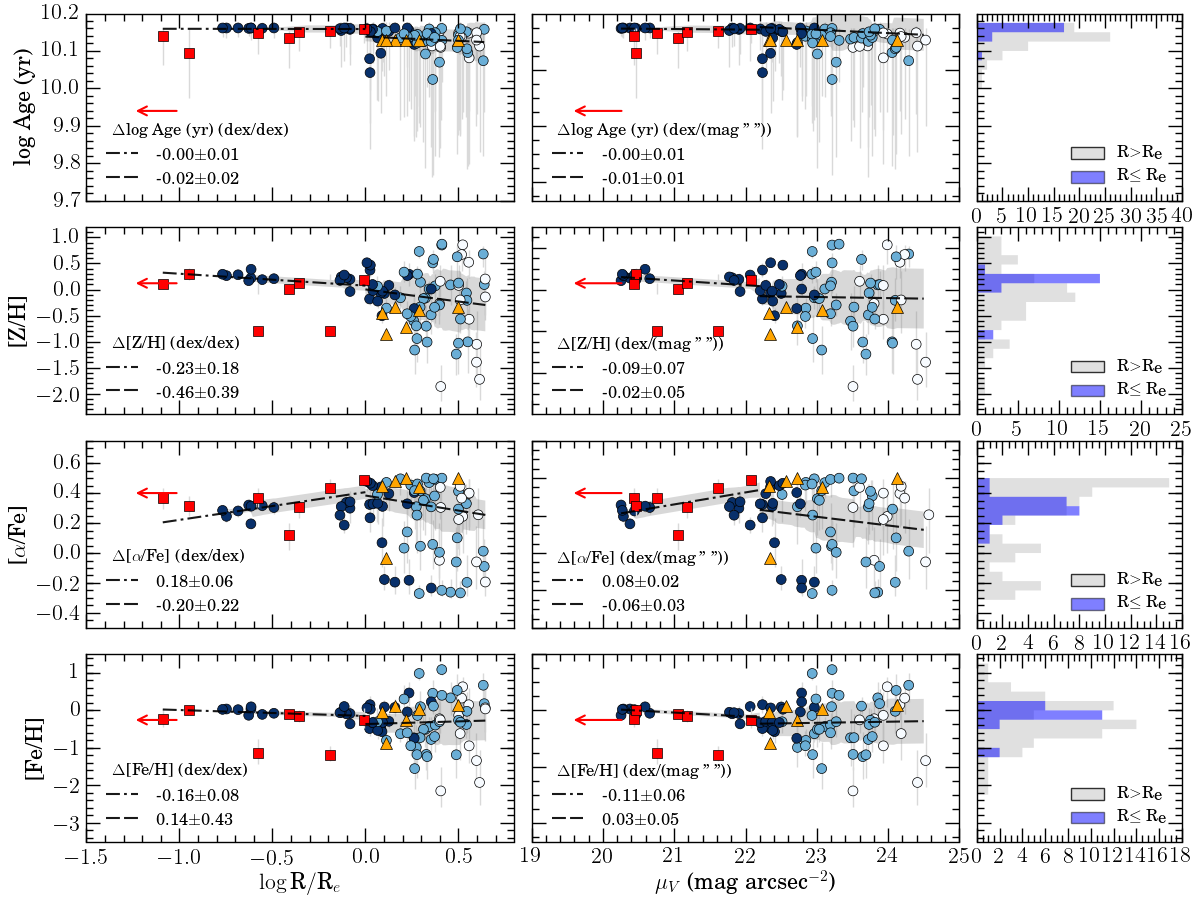}
\caption{Gradients of the stellar population parameters. {Left:} Ages, [Z/H], [$\alpha$/Fe] and [Fe/H] as function of the log-distance to the centre of NGC 3311. Circles indicate data points from this work, coloured according to their signal-to-noise as dark blue (S/N$>30$), light blue ($15\leq$S/N$\leq30$) and white (S/N$<15$). Data from \citet{2009MNRAS.398..133L} and Coccato et al. (2011) are shown by the red arrow and orange triangles, respectively. New calculated values for the stellar population parameters based on data from \citet{2012MNRAS.425..841L} are indicated by red squares. Dot-dashed and dashed lines indicate the regression for the inner and external halo, respectively, with the gradients displayed in the bottom left of each panel. The gray shades represent the systematic error of under/over subtraction of the sky by 1\%. {Centre:} Same as for the left panels, but with gradients measured as a function of the $V$-band surface brightness. {Right:} Histograms of the distribution of the stellar population parameters, combining our data with those of \citet{2012MNRAS.425..841L}, with the inner galaxy and external halo shown in blue and grey bins  respectively. {The gradients in the inner galaxy are consistent with the predictions from a quasi-monolithic collapse model, while the shallow azimuthally-averaged gradients and the large scatter in the outer halo are consistent with the results of accreted stars from a variety of different progenitors.}}
\label{fig:ssp_gradients}
\end{figure*}

To describe the large-scale distribution, we again recur to the analysis of the radial trends and their gradients. In Fig.~\ref{fig:ssp_gradients}, we plot the stellar population parameter values determined from this work, in circles with different grades of blue according to their S/N, as function of both the radial distance from the centre of NGC~3311 and also as a function of the $V$-band surface brightness at the location of the slits (left and central panels, respectively). We also show the stellar population parameters derived from the observations of \citet{2012MNRAS.425..841L} and \citet{2011MNRAS.412.2183T}, as red squares and orange triangles respectively, calculated from their published values of the Lick indices, using our MCMC approach for consistency. We also show the central stellar population parameters from \citet{2009MNRAS.398..133L} using a red arrow indicating the published value in their paper. 

There is a considerable scatter in the external halo properties, similarly to that of the Lick indices plotted in Fig.~\ref{fig:radindices}. Once more, we compare this scatter with the systematic effect of an incorrect sky subtraction by $\pm1$\%, as displayed by the grey shaded areas. The comparison shows that the observed scatter is larger than that caused by a systematic effect in the sky subtraction of $\pm 1\%$. In order to reproduce a scatter similar to the one measured in the external halo, an error of $\sim 6$\% is necessary, which is way larger than any of the residuals in our extracted spectra. Therefore, we have evidence that the scatter at large radii is an intrinsic property of the external stellar halo.  

Similarly to the method deployed for the Lick indices, we quantify the radial trends in the stellar population properties using gradients which are computed separately for the inner galaxy and the external halo. The separation between these two regions is set at the projected distance of $1 R_e$ from the centre, which is approximately represented by the isophotal surface brightness level of $\mu_v\approx 22.2$ mag arcsec$^{\rm{-2}}$. We used equation~\eqref{eq:gradients} to parametrize the gradients; this time we did use two parameters for the abscissa, $X=\{\log ({R/R_e}), \mu_v\}$. For this calculation of the gradients, we excluded the slits around NGC 3309 and the three slits covering the dwarf galaxy HCC~007 at the south of NGC~3311 to avoid contamination. One important remark is that the stellar population gradients as function of the radius from the centre of NGC~3311 is our primary diagnostic to characterize the global variations in stellar population parameters, and as such, is going to be discussed in detail in the following. We expect the gradients as function of the local surface brightness values to be affected by the presence of substructures, as it is the case for the external halo of NGC~3311 \citep[see][]{2012A&A...545A..37A}, and hence have a larger scatter than radial gradients.

\begin{table*}
  \caption{{Linear regression coefficients for the fitting of the stellar population parameters}}         
\label{tab:spgrad1}   
\centering                         
\begin{tabular}{cx{3.5cm}x{3.5cm}x{3.5cm}x{3.5cm}}        
\hline\hline                 
 &  \multicolumn{2}{x{7cm}}{Inner galaxy (1)} & \multicolumn{2}{x{7cm}}{External halo (2)}\\
Property       & $I(0)$ & $\Delta I$ & $I(0)$ & $\Delta I$ \\
 & (dex)  & (dex / dex)& (dex) & (dex / dex)\\ 
\hline                        
log Age (yr) & $10.16\pm0.01$ & $-0.00\pm0.01$ & $10.14\pm0.01$ & $-0.02\pm0.02$\\

[Z/H] & $0.07\pm0.11$ & $-0.23\pm0.18$ & $0.00\pm0.15$ & $-0.46\pm0.39$\\

[$\alpha$/Fe] & $0.41\pm0.03$ & $0.18\pm0.06$ & $0.38\pm0.08$ & $-0.20\pm0.22$\\

[Fe/H] & $-0.15\pm0.08$ & $-0.16\pm0.08$ & $-0.36\pm0.16$ & $0.14\pm0.43$\\
\hline  \hline                                 
\end{tabular}
\tablefoot{ (1) Linear regression coefficients for the fitting of the stellar population parameters in the inner galaxy ($R\leq R_e$) according to equation \eqref{eq:gradients}, parametrized by the logarithm of the radius to the centre of NGC~3311. $I(0)$ indicates the value of the stellar population properties at one effective radius, while $\Delta I$ indicates the gradient. (2) Same as (1) for the outer halo ($R> R_e$) region.}
\end{table*}

\begin{table*}
\caption{{Linear regression coefficients as presented in Table \ref{tab:spgrad1}, parametrized by the V-band surface brightness.}}             
\label{tab:spgrad2}   
\centering                         
\begin{tabular}{cx{3.5cm}x{3.5cm}x{3.5cm}x{3.5cm}}        
\hline\hline                 
 &  \multicolumn{2}{x{7cm}}{Inner galaxy ($\mu_V\leq 22.2$ mag arcsec$^{\rm{-2}}$)} & \multicolumn{2}{x{7cm}}{External halo ($\mu_V> 22.2$ mag arcsec$^{\rm{-2}}$)}\\
 Property      & $I(22.2)$ & $\Delta I$ & $I(22.2)$ & $\Delta I$ \\
 &(dex) & (dex / (mag arcsec$^{-2}$)) & (dex) & (dex (mag arcsec$^{-2}$))\\ 
\hline 
log Age (yr) & $10.19\pm0.05$ & $-0.00\pm0.01$ & $10.36\pm0.11$ & $-0.01\pm0.01$\\

[Z/H] & $2.06\pm1.50$ & $-0.09\pm0.07$ & $0.33\pm1.20$ & $-0.02\pm0.05$\\

[$\alpha$/Fe] & $-1.36\pm0.47$ & $0.08\pm0.02$ & $1.58\pm0.62$ & $-0.06\pm0.03$\\

[Fe/H] & $2.16\pm1.25$ & $-0.11\pm0.06$ & $-1.01\pm1.27$ & $0.03\pm0.05$\\
    
\hline  \hline                                 
\end{tabular}
\end{table*}

\section{Correspondence among stellar populations, surface brightness components and kinematical structures in the inner galaxy and external halo of NGC~3311}
\label{sec:correspondence}

The presence of a break in the measured Lick indices and the clearly distinct distribution of the stellar population parameters for the inner galaxy ($R< R_e$) and the external halo point towards different formation channels for the stars in these two regions. We derive consistent values for the ages, the total metallicity and gradients in [$\alpha$/Fe] and [Fe/H] with very small scatter for the inner galaxy. The distribution of values derived for the external halo have a very large scatter in comparison. The findings for the inner galaxy are consistent with the expectations from the \textit{in situ} formation, that maintains gradients from the time of initial rapid star formation. Differently, the larger scatter of the values in the external halo suggests that these stars come from the debris of gravitationally disrupted galaxies, as suggested by classical model \citep{1979ApJ...231..659D} for the formation of cD halos. {Thus the breaks at $1R_e$ are interpreted as transition from the \textit{in situ} to the accreted stellar populations, and are observed in other giant ellipticals from the MASSIVE survey \citep{2013ApJ...776...64G,2015ApJ...807...11G}.} 

{In NGC~3311, such a break in the stellar properties of the inner galaxy and the external halo correlates with variations of the velocity dispersion profile}. Within one $R_e$, stars move under the influence of the galaxy mass, reaching a $\sigma_0\approx 160$ km s$^{\rm{-1}}$ at the centre. At one $R_e$ and slightly larger radii, the line-of-sight (LOS) velocity dispersion has a positive gradient reaching $\sigma(R) \approx 400$ km s$^{\rm{-1}}$ at 30 kpc \citep[\citetalias{hilker2015}]{2010A&A...520L...9V,2011A&A...531A.119R}. Such an increase in $\sigma(R)$ with radius indicates that the stars are progressively driven by the massive external halo associated with the Hydra I cluster, as mapped by the hot X-ray emission \citep{2004PASJ...56..743H}.  

\begin{figure*}
\centering
\includegraphics[width=\linewidth]{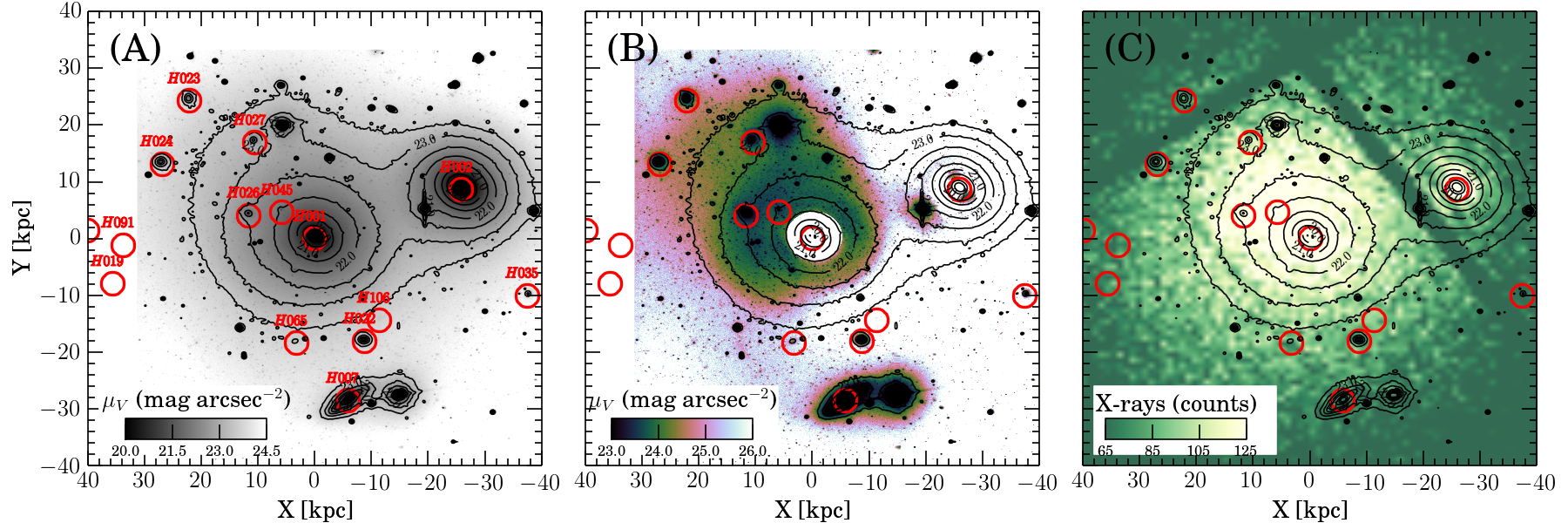}
\caption{Evidence for a large-scale component in the external halo of NGC~3311. (A-B) $V$-band image and residual from the S\'ersic $n=10$ model from \citet{2012A&A...545A..37A} illustration the presence of an off-centred envelope. (C) XMM-Newton X-rays image from \citet{2006PASJ...58..695H} indicating an excess emission at the same position of the off-centred envelope. Red circles indicate the position of dwarf galaxies from the catalog of \citet{2008A&A...486..697M}.}
\label{fig:allmaps}
\end{figure*}

We now discuss the properties of the external halo of NGC~3311, defined as the region $R> R_e$. This halo is not homogeneous, and the presence of additional components can be physically motivated. Using deep $V$-band imaging, \citet{2012A&A...545A..37A} showed that the light of NGC~3311 is described primarily by a single S\'ersic function with index $n\approx 10$ at all radii. There is no break or no photometric signature of the two-component structure expected in a cD galaxy: the cluster dominated halo is not obvious in the photometry. This S\'ersic $n=10$ component centred on NGC~3311 fits most of the light at all directions and is responsible for the featureless appearance observed in the panel A of Fig.~\ref{fig:allmaps}. However, once the symmetric main component is subtracted off, \citet{2012A&A...545A..37A} detected an additional feature in the galaxy light, the off-centred envelope, that is located at a projected distance of $~18$ kpc from the centre of NGC~3311, towards the North-East direction (panel B of Fig.~\ref{fig:allmaps}). At the location of the off-centred envelope, there is an excess emission in X-rays \citep[][see panel C]{2004PASJ...56..743H,2006PASJ...58..695H} and a group of dwarf galaxies are found around this position, with high relative LOS velocities of $\sim 1000$ km s$^{\rm{-1}}$ \citep{2008A&A...486..697M} with respect of the systemic velocity of the Hydra I cluster. At the optical wavelengths, the off-centred envelope is fainter than the symmetric main halo, contributing up to a fraction of $\approx 30$\% of the light \citep{2011A&A...533A.138C,2012A&A...545A..37A}. 

The off-centred envelope is also associated with kinematical signatures. \citet{2011A&A...528A..24V} showed that the line-of-sight velocity distribution (LOSVD) of planetary nebulae in the cluster core around NGC~3311 halo is multi-peaked, with three distinct components: a broad asymmetric component with velocities of $\sim3100$ km s$^{\rm -1}$, a blue-shifted north-south elongated component at $\sim1800$ km s$^{\rm -1}$ and a red-shifted component at $\sim5000$ km s$^{\rm -1}$, at the location of the off-centred envelope. Furthermore \citet{2012A&A...545A..37A} showed that asymmetric features in the velocity dispersion and the LOS velocity profiles correlate with the spatial location of the off-centred envelope. That is, at the location of the envelope, the LOS velocities (LOSVs) are redshifted with respect to the centre of NGC~3311 and the velocity dispersion is larger than at the symmetric location in the external halo, opposite the galaxy centre. Such features can be explained by the superposition along the LOS of two distinct structural components, the S\'ersic n=10 halo and the off-centred envelope, with different LOSVs by $\ge 50$ kms$^{-1}$. 

Therefore, the external halo properties may be associated with two structural components: the ``symmetric'' S\'ersic $n=10$ halo, which is found at all azimuthal angles, and the off-centred envelope in the North East quadrant. The exact boundaries of these components are not well defined, hence we adopt a simple scheme to seek for their signatures in the distribution of the parameters for the stellar populations. The properties of the stellar population in the symmetric external halo are derived for $R > R_e$ and $90\degree\le\mbox{PA}\le360\degree$, while the population of the off-centred halo is studied in the quadrant at $0\degree<\mbox{PA}<90\degree$. As the symmetric halo contributes $70\%$ or even larger fractions to the light at the location of the off-centred envelope, we would expect to only detect small variations in the distributions of the stellar population parameters at the location of the off-centred envelope with respect to the symmetric halo distribution. 

\section{Properties of the stellar populations of the stellar light in NGC~3311}
\label{sec:discussion}

\subsection{Stellar populations in the inner galaxy}
 
The stellar population properties of the inner galaxy ($R\lesssim R_e$) have well constrained values as indicated by the blue histograms on the right side of Fig.~\ref{fig:ssp_gradients}. The radial linear gradients of the stellar parameters are minor. There are three deviating data points: one at the position of the dust lane in the central kpc of NGC~3311 \citep{2012A&A...545A..37A}, and two in other regions further out, see \citet{2009MNRAS.398..133L} and \citet{2012MNRAS.425..841L}, 

The stars of the inner galaxy are old, compatible with the oldest modelled stellar populations ($15$ Gyr), with no gradient in age. The total metallicity of the stars is super solar, with a mild gradient of $-0.23\pm0.18$ dex dex$^{\rm{-1}}$, while the alpha element abundance is high at the centre, [$\alpha$/Fe]$\approx 0.2$ dex, with a positive gradient of $0.18\pm0.06$ dex dex$^{\rm{-1}}$. The resulting iron abundance is close to solar with a gradient of $-0.16\pm0.08$ dex dex$^{\rm{-1}}$. 

The very old age of NGC~3311 is expected for BCGs. \citet{2015MNRAS.449.3347O} observed that one-third of their BCG sample has similarly old stellar populations (age $>12$ Gyr), while \citet{2009MNRAS.398..133L} found that about 50\% of the BCGs have central old stellar populations. The flat age gradient and the abundance gradients are consistent with the values observed in BCGs by \citet{2015MNRAS.449.3347O}. 

\subsection{Stellar populations in the external halo}

The properties of the outer stellar halo of NGC 3311 are considerably different from those inferred for the inner galaxy, as shown in Fig.~\ref{fig:ssp_gradients}, left and central panels at $R\gtrsim R_e$ and $\mu_v \gtrsim 22.2$ mag arcsec$^{\rm{-2}}$. We detect a shallow age gradient of $\Delta \log \mbox{Age}=-0.02\pm0.02$ dex dex$^{{\rm -1}}$ in the external halo, which is compatible with the flat gradient in the inner galaxy. The total metallicity gradient is steeper than that of the inner galaxy, with $\Delta \mbox{[Z/H]}=-0.46\pm0.39$ dex dex$^{{\rm -1}}$, and an inversion of the radial trends is observed in the alpha element abundance, with $\Delta \mbox{[$\alpha$/Fe]}=-0.20\pm0.22$ dex dex$^{{\rm -1}}$, and in the abundance of iron, with $\Delta \mbox{[Fe/H]}=0.14\pm0.43$ dex dex$^{{\rm -1}}$. 

In addition to the average radial trends, the clear feature of the outer stellar halo of NGC~3311 is the considerable larger scatter of the stellar population parameters. The histograms on the right side of Fig.~\ref{fig:ssp_gradients}, show the distribution of stellar population properties for the inner galaxy (blue) and outer stellar halo (gray). The large scatter is present for the SSP parameters of high S/N ($\geq 40$) spectra and at a radius of $R\sim 0.6 R_e = 5.3$ kpc, as indicated by the dark blue circles in the radial plots for [Z/H], [$\alpha$/Fe] and [Fe/H] in Fig.~\ref{fig:ssp_gradients}. Moreover, the width of the distribution of the stellar population parameters is larger than what is expected from a systematic error in the sky subtraction of $\pm 1$\%, as indicated by the grey shaded areas shown around the mean gradients. 

In the histograms on the right side of Fig.~\ref{fig:ssp_gradients}, the positions of the mean peak of the distributions and their widths are different in the inner galaxy and external halo. A common characteristic of these distributions is that their widths for the external halo are twice as large as those for the inner galaxy. In the case of the [Z/H] and [$\alpha$/Fe] distributions, there are multiple peaks for the external halo. Therefore, contrary to the inner galaxy, which can be explained by a rapid process of collapse and merger of gas-rich lumps, the outer stellar halo was most likely built up by accretion of stars from a variety of progenitors, with different masses and star formation histories. 

\begin{figure}
\centering
\includegraphics[width=\linewidth]{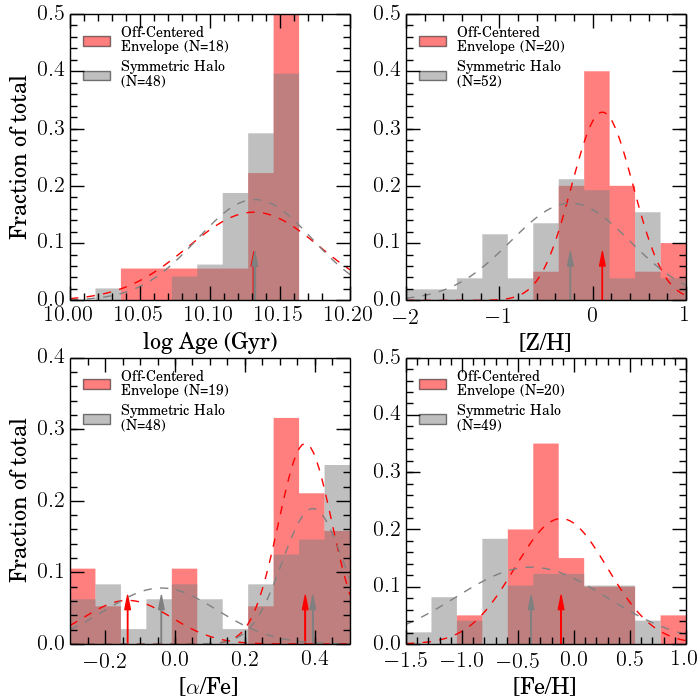}
\caption{Distribution of stellar population parameters of the off-centred envelope (red) and the symmetric halo (gray). The histograms are normalized, and the number of spectra (N) used in each case is shown in the legend of each panel. Dashed lines indicate the components of the GMM analysis, while the arrows indicate the position of the peak of each component. {The off-centred envelope is responsible for a significant shift of the metallicity where it is located, while ages and alpha-element abundance distributions are similar to those inferred for the symmetric halo.}}
\label{fig:outerhist}
\end{figure}

We now investigate whether there are differences in the stellar population parameters between the off-centred envelope and the rest of the external halo of NGC~3311. Fig.~\ref{fig:outerhist} shows the distribution of the stellar population  parameters for the off-centred envelope (red) and the rest of the external halo (grey). To determine the number of peaks that are statistically significant in each case, we have used Gaussian Mixture Models (GMM) to estimate the number of populations which minimize the Bayesian Information Criteria (BIC), whose results are summarized in Table~\ref{tab:outerstats}. The [$\alpha$/Fe] distributions for the off-centred envelope and symmetric halo are those for which two components are statistically favoured with respect to the single peak distribution. The distributions for the ages and metallicities can be relatively well described by single Gaussian functions instead. The GMM models are shown in Fig.~\ref{fig:outerhist} by dashed lines, with arrows indicating the position of the peak for each component. 

 \begin{table}
 \caption{Properties of the Gaussian Mixture Models with lowest Bayesian Information Criteria (BIC) for the off-centred envelope and the symmetric halo.}             
 \label{tab:outerstats}   
 \centering                         
 \begin{tabular}{cccccc}        
 \hline   \hline             
Property & Halo & N & Mean & Sigma & Weight\\ 
(1)      &  (2) &(3)&  (4) &   (5) &   (6) \\
\hline 
log Age & Envelope & 18 & 10.13 & 0.05 & 1.0\\ 

    (Gyr)                            & Sym. Halo & 48 & 10.13 & 0.04 & 1.0\\
                                
\multicolumn{6}{c}{- - - - - -}\\

\multirow{2}{*}{[Z/H]} & Envelope & 20 & 0.10 & 0.33 & 1.0\\

                       & Sym. Halo & 52 & -0.24 & 0.64 & 1.0\\
                       
\multicolumn{6}{c}{- - - - - -}\\

\multirow{4}{*}{[$\alpha$/Fe]} &  Envelope & 19 &  0.37 & 0.08 & 0.74\\

                               &           &    & -0.14 & 0.13 & 0.26\\
  
                               & Sym. Halo & 48 & 0.39 & 0.09 & 0.57\\
                               &           &     & -0.04 & 0.16 & 0.43\\
\multicolumn{6}{c}{- - - - - -}\\

\multirow{2}{*}{[Fe/H]} &  Envelope & 20 & -0.12 & 0.41 & 1.0\\

 & Sym. Halo & 49 & -0.38 & 0.68 & 1.0\\
 \hline                                   
 \end{tabular}
 \tablefoot{{(1) Stellar population parameters. (2) External halo component ($R>R_e$), defined as envelope for the NE quadrant of our observation and symmetric halo otherwise. (3) Number of data points used in the statistics. (4-5) Mean and standard deviation of the Gaussian Mixture Model component. (6) Weight of the component in the Gaussian Mixture Model.}}
 \end{table}

{We used the Kolmogorov-Smirnov (KS) test to statistically compare the distributions of the stellar population parameters for the off-centred envelope and the symmetric halo. The result of the KS tests indicates that the age and the alpha element abundances of these two  components may have been drawn from the same parent distribution, with probabilities of 78 and 32\% respectively.} In both components, stars are predominantly old, with ages of $\sim$14 Gyr, compatible with the oldest stellar populations in the models of \citet{2011MNRAS.412.2183T}. This old age distribution extends the results from \citet{2011A&A...533A.138C} to the entire halo, and is also supported by the lack of features in the UV emission from GALEX observations \citep{2007ApJS..173..185G}. The age is the only external halo property for which the gradient is representative of the vast majority of the observed data points. 

The distribution of alpha element abundance in the symmetric halo and in the off-centred envelope show similar multi-modal distributions, modelled as a mixture of two Gaussian components. The main contribution to the alpha element abundance are stars with [$\alpha$/Fe]$\approx 0.4$ dex and spread of 0.1 dex, observed in 64\% of the cases, while the secondary population has an abundance of [$\alpha$/Fe]$\approx -0.1$ dex and spread of 0.15 dex. The explanation for almost identical alpha element abundance at the off-centred envelope and in the symmetric halo is that this property is robust against a contamination of a minor component to the total integrated light, as indicated by the simulations of  \citet{2011A&A...533A.138C}. Comparing the SSP parameters of a superposition of a typical, alpha-enhanced symmetric halo spectrum ([Z/H]$=-$0.34 dex, [$\alpha$/Fe]=0.50 dex) with the properties of the dwarf galaxy HCC 026 ([Z/H]$=-$0.85 dex, [$\alpha$/Fe]=0.03 dex), they have noticed that a fraction of 35\% in the light is necessary to decrease the element abundance by only 0.05 dex. There may be specific regions in the off-centred envelope where this fraction may be slightly larger, especially at those regions with the largest velocity dispersion observed in \citetalias{hilker2015} but, considering the total light, the fraction of light at the sub structure should not exceed 30\%, confirming the findings of  \citet{2011A&A...533A.138C} and in agreement with the estimates of \citet{2012A&A...545A..37A}. 

Differently from the age and [$\alpha$/Fe] distributions, the total metallicity and the iron abundance distributions do differ between the off-centred envelope and the symmetric halo. The probability that these distributions are drawn from the same parent distributions are low, i.e.  $\sim $1\% and $\sim 3$\%, respectively. The total metallicity distribution of the off-centred envelope is most discrepant. The difference is clearly shown  in Fig.~\ref{fig:outerhist}: the metallicity distribution of the stellar populations at the location of the off-set envelope is a single peaked distribution centred at  [Z/H]$=0.1$ dex and width of $0.33$ dex, while the metallicity of the symmetric halo is a much broader distribution, with mean value  [Z/H]$=-0.24$ dex and width of $0.64$ dex. The stars in the quadrant where the off-centred envelope is located have higher metallicities than most of the stars at the same radii, except for the stars sampled by spectra extracted in proximity of the giant companion NGC~3309. A similar result holds for the iron abundance, with the symmetric halo having [Fe/H]$=-0.38$ dex and dispersion of $0.68$ dex, while the mean abundances of the off-centred envelope is [Fe/H]$=-0.12$ with a dispersion of $0.41$ dex. Since the light at the location of the off-centred envelope is given by the superposition of this component and the symmetric halo, the intrinsic metallicity of the stars in the envelope may be significantly higher than solar. 

Our derived stellar population parameters for the external halo of NGC~3311 are compatible with the previous results from \citet{2011A&A...533A.138C}, and are similar to the stellar populations of other cD galaxy halos, such as NGC~4889 in the Coma cluster \citep{2010MNRAS.407L..26C} and NGC~6166 in the Abell 2199 cluster \citep{2015ApJ...807...56B}. These physical properties are also similar to those inferred for the stellar populations in other BCGs \citep{2008MNRAS.385..675S} and massive early-type galaxies \citep{2010MNRAS.408..272S,2012ApJ...750...32G,2013ApJ...776...64G,2015ApJ...807...11G}.

\section{Implications for the assembly history of NGC~3311}
\label{sec:implication}

Recent cosmological models for the formation of large early-type galaxies predict that the majority of stars in the external halos originated from satellite galaxies in the so-called two-phase scenario \citep[e.g.][]{2007MNRAS.375....2D,2009ApJ...699L.178N,2010ApJ...725.2312O,2013MNRAS.434.3348C}. In the next sections, we are going to compare our results with recent modelling of BCGs in order to relate the inferred distributions of the stellar population parameters with the physical mechanisms involved in the formation of extended halos of BCGs in clusters.  

\subsection{The inner galaxy}

As presented in the previous sections, the distribution and the distinct radial gradients of the age, [Z/H], [Fe/H] and [$\alpha$/Fe] of the stellar populations in the inner galaxy are consistent with stars born \textit{in situ}. According to models of \citet{2013MNRAS.434.3348C}, the large fraction of \textit{in situ} stars  in the central regions of BCGs is expected to occur up to a stellar mass threshold of $M_{200}\lesssim10^{13}M_\odot$, as for more massive galaxies the accreted stars become increasingly dominant even in their innermost regions \citep{2015MNRAS.451.2703C}. NGC~3311 is very close to this boundary, with $M_{200}\approx 0.63 M_{500} = 6.3 \cdot 10^{13} M_\odot$ \citep{2011A&A...534A.109P}, and our determination of the stellar population parameters show the presence of \textit{in situ} and accreted stars in two radial regions, hence in agreement with the predictions of \citet{2013MNRAS.434.3348C}. 

We can also compare our results with those for other massive early-type galaxies. The total metallicity gradient of the inner galaxy in NGC~3311, $\Delta $[Z/H]$=-0.23\pm0.18$ dex dex$^{\rm{-1}}$, is comparable to the gradients of other BCG galaxies \citep{2015MNRAS.449.3347O} and non-BCGs \citep{2010MNRAS.408...97K}. Positive [$\alpha$/Fe] gradients are also common among early-type galaxies \citep{2010MNRAS.408...97K} {and BCGs \citep{2007MNRAS.378.1507B}} and, therefore, the mechanisms that set the gradients in BCGs must be similar to those that form ordinary early-type galaxies. In a simple, quasi-monolithic scenario, such gradients can be explained in an outside-in scenario \citep[e.g.,][]{2006ApJ...638..739P,2008A&A...484..679P,2010MNRAS.407.1347P}. In this context, early-type galaxies are formed by the merging of gas-rich lumps which produces an intense star formation, but with a differential rate at different radii. Star formation in the outer regions around 1$R_e$ is characterized by short time-scales and strong stellar winds that deplete the iron efficiently, whereas star formation is the core is prolonged and metals, in particular iron, are kept due to the strong gravitational potential. This produces positive alpha element gradients and negative total metallicity gradients. The abundance gradients of the inner galaxy of NGC~3311 are compatible with the results of such models considering a few episodes of dry mergers \citep{2010MNRAS.407.1347P}. 

\subsection{The extended symmetric halo}

With reference to the main peak of the alpha element abundance distribution, see Fig.~\ref{fig:outerhist}, the extended halo can be described as a mix of old stars, with metallicities of [Z/H]$\approx -0.25$ and [$\alpha$/Fe]$\approx 0.4$, at first order. As discussed earlier, the high value for the $\alpha$ abundance does not imply that all stars were formed in galaxies with high [$\alpha$/Fe], as this quantity does not change a lot by the contamination of  less alpha enhanced stars. Still the majority of the stars in the external halo  may have indeed high [$\alpha$/Fe]. In the nearby universe, high [$\alpha$/Fe] stars are those:
\begin{enumerate}
  \item Stars produced in disky galaxies with truncated star formation. \citet{2012ApJ...750...32G,2013ApJ...776...64G} have shown that the metallicity and the alpha element abundance at the outskirts of massive galaxies are similar to those found in the Milk-Way thick disk, which is predicted to have been formed in short time-scales \citep{1997ApJ...477..765C}. These disks could have been destroyed in interactions and mergers very early, as galaxy encounters at high redshift were much more common than at the present day universe.
  \item Stars from galaxies in compact groups. \citet{2007AJ....133..330D} have found that galaxies with $\sigma_0 \lesssim 160$ km s$^{-1}$ in Hickson Compact Groups have larger [Mg/Fe] and lower [Z/H] compared to field galaxies of similar masses, and proposed that such galaxies may have their otherwise extended star formation quenched by mergers. 
  \item Stars in the outskirts of massive early-type galaxies with declining metallicity gradients and flat-to-positive alpha element gradients \citep{2011A&A...533A.138C}. Negative metallicity gradients extending to the outer radii are found in most early-type galaxies \citep[e.g.][]{2007A&A...467..991B,2012MNRAS.426.2300L}. Flat and positive  [$\alpha$/Fe] gradients are commonly found in early-type galaxies at small radii \citep{2010MNRAS.408...97K}, and there is evidence that at least metallicity gradients extend to larger radii \citep{2012MNRAS.426.2300L,2014MNRAS.442.1003P}. In other nearby clusters, such as Virgo and Fornax, declining metallicity gradients and flat alpha element abundances are found in almost all cases  \citep{2010MNRAS.408..272S}, but there is no such information for the giant ellipticals in the Hydra I cluster. If those have similar gradients, the central stellar populations of the Hydra I ellipticals should exhibit high [Z/H] and [$\alpha$/Fe], similar to those in the extended halo of NGC 3311 \citep{2011A&A...533A.138C}. 
\end{enumerate}

All of the above galaxies could have provided stars that are now found to contribute most of the light of the  symmetric external halo. They do not exclude each other necessarily, as they could have been different parts of the same process in a hierarchical scenario.  More importantly, however, is that the high [$\alpha$/Fe] indicates star formation happening on short time-scales of $\approx 0.1$ Gyr, according to the approximation of \citet{2005ApJ...621..673T}. We can also estimate the most likely masses of the progenitors of the stars in the external halo from the metallicity distribution in Fig.~\ref{fig:outerhist}.  Translating these metallicities into stellar masses according to the relations from \citet{2005MNRAS.362...41G}, the typical galaxy contributing to the formation of the extended symmetric halo are Milky Way-like galaxies, including mostly galaxies in the range $\sim 10^{10}M_\odot$ to $\sim 10^{12}M_\odot$.

The extrapolation of the inner galaxy gradient in [$\alpha$/Fe] matches the position of the dominant alpha element abundance in the external halo, indicating that these stars may have similar origins. Cosmological hydrodynamical simulations from \citet{2007MNRAS.377....2M} indeed indicate that most of the diffuse halo is composed of stars liberated from the most massive galaxies in episodes of major mergers related to the formation of the cD galaxy, where tidal shocking and stripping of massive galaxies are able to unbound up to 30\% of their stellar mass. In agreement with several predictions from the models of \citet{2007MNRAS.377....2M}, semi-analytical models of \citet{2014MNRAS.437.3787C}, including disruption of galaxies and tidal interactions, are able to reproduce both the typical masses of the progenitors in our observations as well as the typical total metallicities. Therefore, our observations are in agreement with models in which the assembly of the BCG is directly connected with the formation of the diffuse halo.

The [$\alpha$/Fe] distribution of the symmetric halo indicates the presence of a secondary population, with solar and sub-solar $\alpha$-abundances. These lower alpha element abundances suggest a much more extended star formation, $\sim$15 Gyr, according to the approximation of \citet{2005ApJ...621..673T}. In this case, there is already one galaxy in the Hydra I cluster which has similar properties, the dwarf galaxy HCC 026. This dwarf galaxy has solar alpha-abundance, [$\alpha$/Fe]$=-0.03\pm0.05$ dex, and low metallicity, [Z/H]=$-0.85\pm0.03$ \citep{2011A&A...533A.138C}. HCC 026 is part of a group of dwarf galaxies at the same location of the off-centred envelope \citep{2008A&A...486..697M} and it has strong tidal tails that indicate that this galaxy is being disrupted  by the tidal field close to the cluster centre \citep{2012A&A...545A..37A}. Hence 
HCC 026 is the typical object that contributes to the late accretion events that build up the halo today. 

Such solar and sub-solar $\alpha$ abundance populations are inferred from a substantial fraction of the extracted spectra, in about 40\% of the total sample in the external halo. Hence accretion of stars from disrupted dwarfs or irregular galaxies is an important channel for the late build-up of the external halo. Very recent observations of the planetary nebulae in the halo of M87 and the intracluster light in the Virgo core showed that the contribution from Magellanic clouds-like irregular galaxies is responsible for a sizeable fraction of the halo light being added in the last Gyr to the M87 halo and ICL \citep{2015A&A...579A.135L,2015A&A...579L...3L}.
Models from \citet{2007MNRAS.377....2M} indicate that low-mass galaxies ($M_*<10^{10}M_\odot$) may contribute to the formation of the diffuse halo, but numerical issues related to the particle resolution of low-mass galaxies did not allow them to set proper constraints on this secondary population. 

Our current results for NGC~3311 and the recent findings for M87 indicate
that it is of importance to study the details of the late mass accretion, which is responsible for a sizeable fraction of the chemical composition and kinematics of the halos of massive nearby galaxies. 

\subsection{The metallicity distribution of the off-centred envelope}

The stellar populations located at the off-centred envelope show an enhancement of metals in comparison with the rest of the symmetric halo, while the alpha element abundance and age distributions are similar. The strongly peaked metallicity distribution centred around [Z/H]$\approx 0.1$ indicates that the stars associated with the substructure were formed in massive galaxies. 
According to \citet{2007MNRAS.377....2M}, massive ellipticals can lose up to 30\% of their stars during their merging events with the cD and these stars would then contribute to the build up of the external halo. One can then ask how this external high metallicity halo component acquired an offset and became off-centred with respect to the high surface brightness, highly concentrated inner galaxy.

Off-centred outer components in BGCs are rather frequent: for a sample of 24 clusters, \citet{2005ApJ...618..195G} showed that a two-component fit to the light profiles of BCGs provides an improved match to the data and that the two photometric components are misaligned in $60\%$  of the sample.  In the \citet{2005ApJ...618..195G} sample, the cluster Abell 1651 has a cD galaxy where the two components have different centres,  with the outer component being off-centred by 15 kpc linear distance in projection, i.e. very similar to what is observed for NGC~3311. As observed in cosmological simulations \citep[e.g.][]{2007MNRAS.377....2M}, the highly radial orbits of the dark halos of massive ellipticals and their tidal interaction with the cD dark halo may cause a deflection of the central part of the cD from its dark halo, while the outer envelopes maintain their orbital directions, thus creating an offset between core and halo in the cD galaxy. Candidate dark halos responsible for a tidal interaction with NGC 3311 are the dark halo associated with the group of infalling dwarfs or the dark halo associated with NGC 3309.  These tidal interactions would then be similar to what is observed in the Coma cluster core \citep{2007A&A...468..815G}. Such  mechanisms may also explain the gas stripping of NGC~3311 as observed in the X-rays \citet{2004PASJ...56..743H,2006PASJ...58..695H}, as the mass of the gas is compatible with the mass of NGC 3311. 

\section{Summary and conclusions}
\label{sec:conclusion}

We performed a spatially extended survey of the stellar populations in NGC~3311, the brightest galaxy of the Hydra I cluster. By the analysis of the absorption line equivalent widths using a Bayesian framework with Monte Carlo Markov Chains, we probed luminosity-weighted parameters -- age, [Z/H], [$\alpha$/Fe] and [Fe/H] -- out to $\sim3R_e$. This enabled the characterisation of the stellar content of three physically motivated components of this system: the inner galaxy, the symmetric external halo and the off-centred envelope. 

The inner stellar halo ($R<R_e$) presents stellar populations typical of massive early-type galaxies, including old ages, high metallicities and high alpha element abundances. Similar to other BCGs, the inner galaxy has a well-defined negative metallicity and  positive alpha element abundance gradient which can be explained by a quasi-monolithic scenario involving a few dry mergers. These gradients and the smaller velocity dispersion of the inner galaxy are clear indications that the stars in this region were formed \textit{in situ} in the early phases of galaxy formation.

The stellar component of the outer symmetric stellar halo is characterized by a large spread of the stellar parameter values rather than by clearly defined radial gradients. This region is characterized also by high velocity dispersions which are indicative that these stars are driven by the cluster's potential and are generated by accumulation of tidal debris. While the mean value of the metallicity distribution in this region indicates sub-solar abundances, the [$\alpha$/Fe] distribution of the symmetric halo is bimodal, with high ($\sim$0.4) and low ($\sim$0) components. The majority of stars in the symmetric halo  are generated in galaxies with a rapid star formation and short time-scales, in agreement with models for the formation of cD envelopes \citep[e.g.][]{2004ApJ...607L..83M,2007MNRAS.377....2M,2014MNRAS.437.3787C}. However, a substantial fraction of stars, about 40\%, has a low [$\alpha$/Fe] value, which indicates that stars from less massive galaxies are also added to the cD halo. Their association with dwarf galaxies currently being disrupted at the position of the off-set envelope indicate that the growth of the cD halo is an on-going process, which is fed by late mass accretion.

Finally, the stellar populations at the position of the off-centred envelope, a substructure also observed in photometry, X-rays and kinematics, indicates that the dark matter halo of an infalling group may have interacted with the BCG halo, causing the stripping of gas and stars. 

We conclude that massive satellite galaxies in the vicinity of NGC 3311 merged with the central cD in early times and formed its symmetric outer halo, while at present time the build up process of the extended halo is still on-going, as indicated by the presence of an infalling group of dwarfs that are adding their stars to the halo now. Tidal interactions between the dark halos of the infalling group, of NGC 3309 and NGC 3311 may be responsible for the stripping of stars and gas in the halo of NGC 3311. Although very challenging from the observational point of view, the halos of cD galaxies provide important constraints on the formation and morphological transformation of galaxies in nearby clusters.  

\begin{acknowledgements}
We would like to thank Dr. Ilani Loubser for kindly sharing data for the inner regions of NGC 3311. We would like to thank the referee for his/her constructive comments. This research has made use of the NASA/IPAC Extragalactic Database (NED) which is operated by the Jet Propulsion Laboratory, California Institute of Technology, under contract with the National Aeronautics and Space Administration. CEB and CMdO are grateful  to the S\~{a}o Paulo Research Foundation (FAPESP) funding (Procs. 2011/21325-0, 2012/22676-3 and 2014/07684-5). TR acknowledges financial support  from FONDECYT project Nr. 1100620, and from the BASAL Centro de Astrof\'{\i}sica y Tecnologias Afines (CATA) PFB-06/2007. TR also thanks ESO/Garching for a visitorship.
\end{acknowledgements}

\bibliographystyle{aa} 
\bibliography{biblio}

\begin{appendix}

\section{Simulations of the variance of Lick indices and stellar populations generated by a systematic error of the sky subtraction at the $\pm 1\%$ level.}

\begin{figure}[!htb]
\centering
\includegraphics[width=0.99\linewidth]{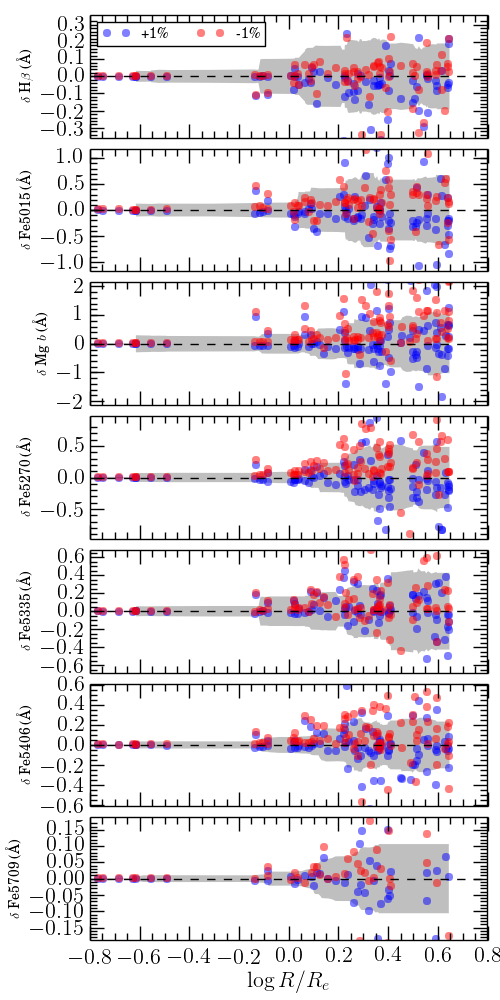}
\caption{Difference of the Lick indices values generated by a systematic error in the sky subtraction of $\pm1\%$ as a function of the distance to the center of NGC 3311. The blue (red) circles represent the measurements performed by the addition (subtraction) of 1\% of the sky spectra. The gray shaded areas represent the rolling standard deviation in the measurements.}
\label{fig:lick_mad}
\end{figure}

\begin{figure}[!htb]
\centering
\includegraphics[width=0.99\linewidth]{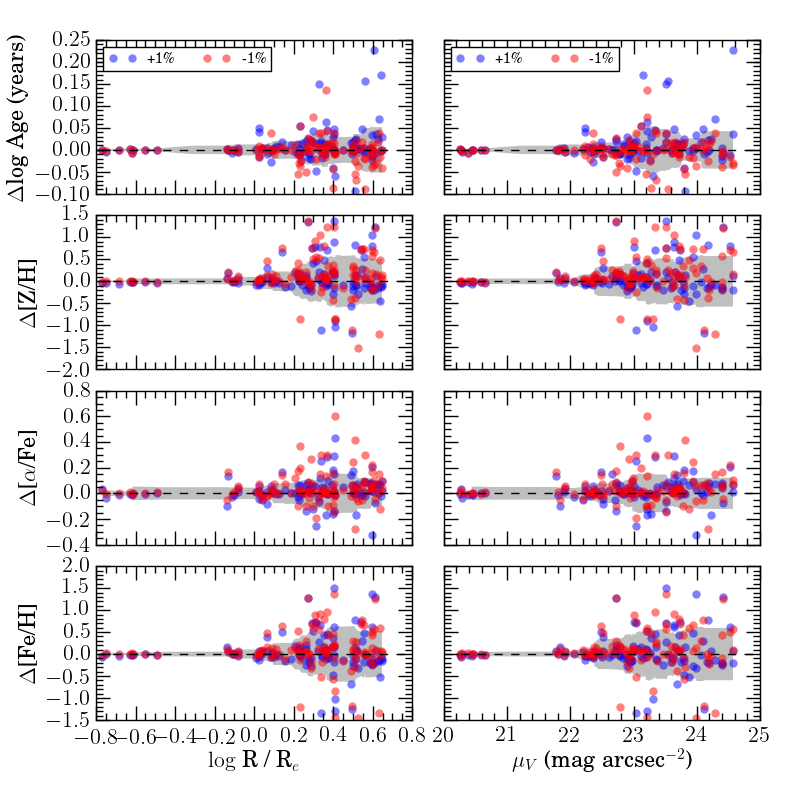}
\caption{Difference of the stellar population parameters generated by a systematic error in the sky subtraction of $\pm1\%$ as a function of the distance to the center of NGC 3311 (left) and as a function of the V-band surface brightness (right). The blue (red) circles represent the measurements performed by the addition (subtraction) of 1\% of the sky spectra. The gray shaded areas represent the rolling standard deviation in the measurements.}
\label{fig:ssps_std}
\end{figure}

\end{appendix}


\end{document}

%% file: lick.tex
cen1 s18 & 19.2 & -131.2 & 15.0 & $2.8\pm1.0$ & $-0.0\pm2.0$ & $1.5\pm1.0$ & -- & $3.5\pm1.2$ & $2.7\pm0.9$ & $3.6\pm0.6$ & $10.11_{-0.35}^{+0.05}$ & $-0.70_{-0.92}^{+0.92}$ & $0.15_{-0.24}^{+0.22}$\\
cen1 s19 & 21.4 & -105.1 & 13.7 & $1.8\pm1.0$ & $-0.0\pm2.0$ & $1.6\pm1.2$ & $1.3\pm1.0$ & $1.9\pm1.2$ & $-2.0\pm0.8$ & -- & $10.11_{-0.19}^{+0.05}$ & $-1.86_{-0.28}^{+0.40}$ & $0.30_{-0.26}^{+0.16}$\\
cen1 s20 & 14.5 & -136.4 & 17.6 & $1.1\pm0.9$ & $6.0\pm2.0$ & $5.7\pm0.9$ & $-0.8\pm0.9$ & $3.5\pm1.0$ & -- & $1.1\pm0.6$ & $10.12_{-0.36}^{+0.04}$ & $-0.68_{-0.90}^{+0.90}$ & $0.16_{-0.25}^{+0.21}$\\
cen1 s21 & 13.0 & -122.9 & 25.2 & $1.3\pm0.6$ & $4.6\pm1.4$ & $4.2\pm0.7$ & $1.9\pm0.7$ & $-1.2\pm0.9$ & $1.3\pm0.8$ & $0.4\pm0.5$ & $10.14_{-0.22}^{+0.03}$ & $-0.16_{-0.22}^{+0.22}$ & $0.50_{-0.11}^{+0.00}$\\
cen1 s23 & 10.6 & -109.8 & 49.8 & $1.1\pm0.3$ & $4.3\pm0.6$ & $3.2\pm0.3$ & $2.8\pm0.3$ & $4.1\pm0.4$ & $1.4\pm0.3$ & $1.3\pm0.2$ & $10.15_{-0.12}^{+0.02}$ & $-0.08_{-0.09}^{+0.11}$ & $-0.18_{-0.07}^{+0.09}$\\
cen1 s24 & 6.9 & -111.5 & 43.2 & $1.1\pm0.4$ & $3.3\pm0.8$ & $4.9\pm0.3$ & $3.8\pm0.4$ & $3.3\pm0.5$ & $1.3\pm0.4$ & $0.7\pm0.2$ & $10.16_{-0.09}^{+0.02}$ & $0.20_{-0.10}^{+0.11}$ & $0.30_{-0.10}^{+0.09}$\\
cen1 s25 & 10.2 & -79.3 & 52.0 & $2.1\pm0.3$ & $0.9\pm0.6$ & $4.0\pm0.3$ & $3.2\pm0.3$ & $1.8\pm0.3$ & $1.6\pm0.3$ & -- & $10.09_{-0.16}^{+0.06}$ & $-0.07_{-0.09}^{+0.11}$ & $0.44_{-0.08}^{+0.05}$\\
cen1 s26 & 6.6 & -76.7 & 69.0 & $0.9\pm0.2$ & $5.2\pm0.5$ & $5.1\pm0.2$ & $2.8\pm0.2$ & $2.5\pm0.3$ & $2.2\pm0.2$ & -- & $10.16_{-0.05}^{+0.01}$ & $0.23_{-0.06}^{+0.07}$ & $0.34_{-0.06}^{+0.06}$\\
cen1 s27 & 2.0 & -91.7 & 114.2 & $1.31\pm0.15$ & $4.7\pm0.3$ & $4.84\pm0.13$ & $3.03\pm0.11$ & $3.11\pm0.12$ & $2.21\pm0.09$ & -- & $10.16_{-0.03}^{+0.01}$ & $0.22_{-0.04}^{+0.04}$ & $0.19_{-0.03}^{+0.03}$\\
cen1 s28 & 10.3 & -52.9 & 44.5 & $1.6\pm0.3$ & $2.6\pm0.8$ & $2.6\pm0.4$ & $1.8\pm0.4$ & $2.2\pm0.4$ & $2.2\pm0.3$ & -- & $10.16_{-0.11}^{+0.02}$ & $-0.50_{-0.15}^{+0.15}$ & $0.07_{-0.16}^{+0.17}$\\
cen1 s29 & 2.3 & 39.9 & 94.1 & $1.0\pm0.2$ & $4.5\pm0.3$ & $5.1\pm0.2$ & $2.92\pm0.14$ & $2.8\pm0.2$ & $2.11\pm0.11$ & $0.8\pm0.09$ & $10.16_{-0.02}^{+0.02}$ & $0.19_{-0.04}^{+0.04}$ & $0.32_{-0.04}^{+0.04}$\\
cen1 s29a & 2.0 & 38.1 & 92.2 & $1.0\pm0.2$ & $4.4\pm0.3$ & $5.1\pm0.2$ & $2.9\pm0.14$ & $2.8\pm0.2$ & $2.1\pm0.12$ & $0.79\pm0.09$ & $10.16_{-0.02}^{+0.02}$ & $0.17_{-0.04}^{+0.04}$ & $0.32_{-0.04}^{+0.04}$\\
cen1 s29b & 2.7 & 41.3 & 85.7 & $1.2\pm0.2$ & $4.3\pm0.4$ & $5.1\pm0.2$ & $3.0\pm0.2$ & $2.7\pm0.2$ & $2.38\pm0.14$ & $0.73\pm0.11$ & $10.16_{-0.03}^{+0.01}$ & $0.21_{-0.04}^{+0.04}$ & $0.30_{-0.04}^{+0.04}$\\
cen1 s30 & 8.9 & -25.5 & 59.4 & $2.0\pm0.2$ & $3.1\pm0.6$ & $3.4\pm0.2$ & $2.4\pm0.3$ & $2.3\pm0.3$ & $1.3\pm0.3$ & -- & $10.08_{-0.17}^{+0.07}$ & $-0.18_{-0.11}^{+0.11}$ & $0.32_{-0.09}^{+0.09}$\\
cen1 s31 & 6.2 & 14.0 & 63.8 & $1.4\pm0.2$ & $5.3\pm0.5$ & $5.0\pm0.2$ & $2.8\pm0.2$ & $3.0\pm0.2$ & $1.8\pm0.2$ & -- & $10.15_{-0.10}^{+0.02}$ & $0.24_{-0.07}^{+0.09}$ & $0.31_{-0.06}^{+0.06}$\\
cen1 s32 & 8.9 & 4.0 & 58.7 & $1.9\pm0.3$ & $6.2\pm0.8$ & $4.8\pm0.3$ & $2.7\pm0.4$ & $2.7\pm0.4$ & $3.2\pm0.4$ & -- & $10.04_{-0.20}^{+0.10}$ & $0.37_{-0.13}^{+0.13}$ & $0.23_{-0.10}^{+0.10}$\\
cen1 s33 & 9.2 & 33.1 & 70.6 & $1.0\pm0.2$ & $4.9\pm0.5$ & $4.4\pm0.2$ & $1.9\pm0.2$ & $2.7\pm0.3$ & $2.0\pm0.3$ & -- & $10.16_{-0.06}^{+0.02}$ & $-0.02_{-0.05}^{+0.06}$ & $0.33_{-0.07}^{+0.07}$\\
cen1 s34 & 11.6 & 25.7 & 40.8 & -- & $3.5\pm0.9$ & $4.3\pm0.4$ & $2.4\pm0.5$ & $2.3\pm0.6$ & $2.1\pm0.6$ & $1.1\pm0.3$ & $10.12_{-0.36}^{+0.04}$ & $-0.66_{-0.91}^{+0.91}$ & $0.10_{-0.23}^{+0.23}$\\
cen1 s35 & 14.0 & 17.3 & 33.9 & $1.0\pm0.5$ & $3.3\pm0.9$ & $4.9\pm0.4$ & $4.1\pm0.5$ & $5.6\pm0.6$ & $1.4\pm0.5$ & -- & $10.16_{-0.00}^{+0.01}$ & $0.47_{-0.12}^{+0.12}$ & $0.01_{-0.11}^{+0.11}$\\
cen1 s36 & 18.4 & 4.7 & 25.1 & -- & $-1.3\pm1.3$ & $3.3\pm0.5$ & $0.5\pm0.7$ & $1.3\pm0.9$ & $2.7\pm0.8$ & -- & $10.11_{-0.36}^{+0.05}$ & $-0.40_{-0.97}^{+0.82}$ & $0.17_{-0.25}^{+0.21}$\\
cen1 s37 & 17.7 & 43.3 & 21.5 & $2.3\pm0.8$ & $4.0\pm2.0$ & $3.9\pm1.0$ & $0.8\pm1.1$ & $2.7\pm1.5$ & $3.0\pm2.0$ & $1.4\pm0.9$ & $10.09_{-0.30}^{+0.07}$ & $-0.17_{-0.36}^{+0.36}$ & $0.36_{-0.24}^{+0.12}$\\
cen2 s21 & 18.5 & 57.8 & 27.0 & $1.7\pm0.5$ & $4.0\pm1.2$ & $4.7\pm0.5$ & $1.9\pm0.5$ & $1.3\pm0.6$ & $3.2\pm0.5$ & $0.4\pm0.3$ & $10.16_{-0.18}^{+0.02}$ & $0.01_{-0.12}^{+0.15}$ & $0.47_{-0.11}^{+0.03}$\\
cen2 s22 & 17.2 & 65.3 & 24.1 & $1.8\pm0.6$ & $5.3\pm1.4$ & $3.4\pm0.6$ & $1.4\pm0.7$ & $3.3\pm0.8$ & $0.2\pm0.8$ & $1.1\pm0.5$ & $10.12_{-0.27}^{+0.04}$ & $-0.17_{-0.23}^{+0.23}$ & $0.35_{-0.20}^{+0.12}$\\
cen2 s23 & 14.2 & 58.7 & 32.6 & $1.0\pm0.5$ & $5.1\pm1.2$ & $4.7\pm0.6$ & $2.5\pm0.5$ & $3.3\pm0.6$ & $1.1\pm0.7$ & $1.0\pm0.4$ & $10.16_{-0.15}^{+0.02}$ & $0.12_{-0.15}^{+0.17}$ & $0.32_{-0.15}^{+0.12}$\\
cen2 s25 & 11.2 & 67.2 & 40.9 & $1.6\pm0.3$ & $4.6\pm0.9$ & $4.8\pm0.4$ & $2.5\pm0.5$ & $3.9\pm0.5$ & $1.1\pm0.4$ & $0.3\pm0.2$ & $10.15_{-0.10}^{+0.02}$ & $0.08_{-0.11}^{+0.11}$ & $0.40_{-0.11}^{+0.08}$\\
cen2 s26 & 10.4 & 85.5 & 35.6 & $0.9\pm0.4$ & $3.2\pm0.9$ & $4.2\pm0.4$ & $2.7\pm0.5$ & $3.2\pm0.6$ & $1.5\pm0.4$ & $0.6\pm0.2$ & $10.16_{-0.09}^{+0.02}$ & $-0.03_{-0.11}^{+0.11}$ & $0.31_{-0.13}^{+0.11}$\\
cen2 s27 & 6.9 & 78.8 & 63.7 & $1.3\pm0.2$ & $5.0\pm0.5$ & $4.3\pm0.2$ & $2.6\pm0.3$ & $2.4\pm0.3$ & $1.4\pm0.2$ & $0.76\pm0.13$ & $10.16_{-0.07}^{+0.02}$ & $-0.04_{-0.06}^{+0.07}$ & $0.34_{-0.07}^{+0.07}$\\
cen2 s28 & 9.8 & 104.5 & 42.4 & $2.0\pm0.3$ & $4.0\pm0.9$ & $3.4\pm0.3$ & $3.1\pm0.5$ & $2.3\pm0.5$ & $1.5\pm0.4$ & $0.0\pm0.2$ & $10.13_{-0.17}^{+0.03}$ & $-0.22_{-0.13}^{+0.13}$ & $0.33_{-0.14}^{+0.11}$\\
cen2 s29 & 6.1 & 104.8 & 68.8 & $0.9\pm0.2$ & $6.5\pm0.5$ & $4.9\pm0.2$ & $2.3\pm0.2$ & $3.1\pm0.3$ & $2.5\pm0.2$ & $0.78\pm0.13$ & $10.16_{-0.05}^{+0.02}$ & $0.18_{-0.06}^{+0.07}$ & $0.25_{-0.06}^{+0.06}$\\
cen2 s30 & 1.5 & 89.0 & 91.2 & $1.1\pm0.2$ & $4.2\pm0.3$ & $5.1\pm0.2$ & $3.34\pm0.15$ & $3.1\pm0.2$ & $2.28\pm0.13$ & $0.96\pm0.09$ & $10.16_{-0.03}^{+0.01}$ & $0.27_{-0.04}^{+0.04}$ & $0.24_{-0.04}^{+0.04}$\\
cen2 s31 & 9.8 & 127.9 & 25.0 & $0.5\pm0.7$ & $5.1\pm1.4$ & $4.6\pm0.8$ & $1.7\pm0.7$ & $3.5\pm1.0$ & $1.3\pm0.9$ & $1.1\pm0.5$ & $10.15_{-0.18}^{+0.02}$ & $0.06_{-0.20}^{+0.20}$ & $0.39_{-0.18}^{+0.09}$\\
cen2 s32 & 1.7 & -138.5 & 93.3 & $1.2\pm0.2$ & $4.0\pm0.3$ & $5.3\pm0.2$ & $3.56\pm0.14$ & $3.0\pm0.2$ & $2.19\pm0.12$ & $0.77\pm0.09$ & $10.16_{-0.02}^{+0.01}$ & $0.28_{-0.04}^{+0.04}$ & $0.28_{-0.04}^{+0.04}$\\
cen2 s32a & 2.0 & -137.2 & 83.1 & $1.1\pm0.2$ & $4.5\pm0.4$ & $5.54\pm0.15$ & $3.4\pm0.2$ & $3.0\pm0.2$ & $2.55\pm0.14$ & $0.82\pm0.1$ & $10.16_{-0.03}^{+0.01}$ & $0.39_{-0.04}^{+0.04}$ & $0.31_{-0.04}^{+0.04}$\\
cen2 s32b & 1.4 & -140.3 & 93.3 & $1.2\pm0.2$ & $4.0\pm0.3$ & $5.26\pm0.13$ & $3.55\pm0.15$ & $3.0\pm0.2$ & $2.19\pm0.13$ & $0.77\pm0.09$ & $10.16_{-0.02}^{+0.01}$ & $0.29_{-0.04}^{+0.04}$ & $0.28_{-0.04}^{+0.04}$\\
cen2 s33 & 9.2 & 146.7 & 36.4 & $1.0\pm0.4$ & $3.9\pm0.9$ & $5.0\pm0.3$ & $2.1\pm0.4$ & $2.9\pm0.5$ & $2.1\pm0.4$ & $0.6\pm0.2$ & $10.16_{-0.09}^{+0.02}$ & $0.10_{-0.09}^{+0.11}$ & $0.45_{-0.08}^{+0.04}$\\
cen2 s34 & 6.5 & 174.5 & 68.5 & $1.1\pm0.2$ & $5.4\pm0.4$ & $4.7\pm0.2$ & $3.5\pm0.2$ & $2.7\pm0.2$ & $2.3\pm0.2$ & -- & $10.16_{-0.09}^{+0.01}$ & $0.28_{-0.06}^{+0.08}$ & $0.19_{-0.05}^{+0.05}$\\
cen2 s35 & 8.9 & 176.0 & 41.0 & $1.1\pm0.4$ & $6.9\pm0.8$ & $5.9\pm0.3$ & $2.2\pm0.4$ & $3.1\pm0.5$ & $2.0\pm0.3$ & $1.4\pm0.3$ & $10.15_{-0.16}^{+0.02}$ & $0.47_{-0.09}^{+0.10}$ & $0.47_{-0.07}^{+0.03}$\\
cen2 s36 & 8.6 & -153.6 & 47.2 & $1.0\pm0.3$ & $3.9\pm0.7$ & $6.3\pm0.3$ & $4.1\pm0.3$ & $2.4\pm0.4$ & $2.4\pm0.3$ & $0.4\pm0.2$ & $10.16_{-0.08}^{+0.02}$ & $0.51_{-0.08}^{+0.08}$ & $0.48_{-0.06}^{+0.02}$\\
cen2 s37 & 10.9 & -164.5 & 29.0 & $1.5\pm0.5$ & $3.6\pm1.1$ & $3.4\pm0.5$ & $1.6\pm0.6$ & $1.1\pm0.7$ & $1.3\pm0.5$ & $0.6\pm0.4$ & $10.15_{-0.17}^{+0.02}$ & $-0.36_{-0.19}^{+0.19}$ & $0.47_{-0.14}^{+0.03}$\\
cen2 s38 & 12.1 & -147.8 & 30.4 & $1.2\pm0.5$ & $2.2\pm1.0$ & $3.2\pm0.4$ & $2.9\pm0.5$ & $4.1\pm0.6$ & $2.8\pm0.4$ & $0.3\pm0.3$ & $10.16_{-0.00}^{+0.01}$ & $-0.10_{-0.13}^{+0.15}$ & $-0.19_{-0.08}^{+0.11}$\\
cen2 s39 & 15.3 & -162.2 & 16.5 & $1.5\pm1.0$ & $3.0\pm3.0$ & $2.3\pm1.3$ & $0.0\pm2.0$ & $4.0\pm2.0$ & $-1.0\pm2.0$ & $2.0\pm1.1$ & $10.13_{-0.27}^{+0.03}$ & $-0.61_{-0.55}^{+0.55}$ & $0.13_{-0.23}^{+0.21}$\\
cen2 s40 & 16.5 & -138.8 & 16.6 & $1.4\pm0.9$ & $1.0\pm2.0$ & $4.1\pm1.1$ & $2.0\pm1.2$ & $3.1\pm1.3$ & $-1.6\pm1.3$ & $1.4\pm0.8$ & $10.14_{-0.24}^{+0.03}$ & $-0.34_{-0.41}^{+0.41}$ & $0.41_{-0.24}^{+0.08}$\\
inn1 s16 & 24.1 & -128.5 & 17.1 & $2.0\pm2.0$ & $11.0\pm2.0$ & $0.2\pm1.0$ & $3.5\pm1.0$ & $4.4\pm1.1$ & $4.4\pm1.0$ & $-0.9\pm0.5$ & $9.62_{-0.34}^{+0.34}$ & $0.09_{-0.39}^{+0.39}$ & $-0.26_{-0.04}^{+0.09}$\\
inn1 s17 & 21.0 & -133.2 & 15.0 & $1.0\pm1.0$ & $5.0\pm3.0$ & $2.9\pm1.3$ & $2.1\pm1.2$ & $7.5\pm1.5$ & $-1.4\pm1.3$ & $-0.4\pm0.8$ & $10.15_{-0.26}^{+0.02}$ & $-0.27_{-0.40}^{+0.40}$ & $-0.05_{-0.17}^{+0.23}$\\
inn1 s18 & 21.4 & -112.8 & 14.7 & -- & $-0.0\pm2.0$ & -- & $6.4\pm1.0$ & $-0.9\pm1.2$ & $3.6\pm0.9$ & $1.2\pm0.7$ & $10.11_{-0.35}^{+0.06}$ & $-0.48_{-0.96}^{+0.85}$ & $0.15_{-0.25}^{+0.22}$\\
inn1 s19 & 17.8 & -120.9 & 20.6 & $3.1\pm0.7$ & $1.0\pm2.0$ & $3.5\pm0.8$ & $1.9\pm0.7$ & $0.8\pm0.9$ & $-1.1\pm0.7$ & $1.3\pm0.5$ & $9.94_{-0.32}^{+0.18}$ & $-0.70_{-0.41}^{+0.41}$ & $0.50_{-0.19}^{+0.00}$\\
inn1 s20 & 21.6 & -96.0 & 27.3 & $-2.9\pm0.5$ & $5.3\pm1.0$ & $4.8\pm0.5$ & $5.2\pm0.5$ & $4.7\pm0.6$ & $4.3\pm0.4$ & $1.0\pm0.4$ & $10.16_{-0.05}^{+0.02}$ & $0.85_{-0.09}^{+0.04}$ & $-0.25_{-0.04}^{+0.07}$\\
inn1 s21 & 16.6 & -102.8 & 28.4 & $2.2\pm0.5$ & $3.9\pm1.2$ & $3.5\pm0.5$ & $5.1\pm0.6$ & $3.2\pm0.7$ & $0.4\pm0.5$ & $0.8\pm0.4$ & $10.04_{-0.34}^{+0.12}$ & $0.14_{-0.21}^{+0.26}$ & $0.05_{-0.16}^{+0.17}$\\
inn1 s23 & 16.2 & -86.2 & 25.0 & $2.1\pm0.6$ & $6.0\pm1.2$ & $4.0\pm0.5$ & $0.8\pm0.6$ & $-0.3\pm0.7$ & $0.9\pm0.5$ & $-1.7\pm0.3$ & $10.11_{-0.36}^{+0.05}$ & $-0.44_{-0.95}^{+0.83}$ & $0.15_{-0.24}^{+0.22}$\\
inn1 s24 & 15.8 & -75.9 & 18.6 & $2.0\pm0.8$ & $3.0\pm2.0$ & $2.5\pm0.8$ & $1.1\pm0.9$ & $1.1\pm1.2$ & $2.0\pm0.9$ & -- & $10.12_{-0.25}^{+0.04}$ & $-0.70_{-0.38}^{+0.38}$ & $0.31_{-0.25}^{+0.15}$\\
inn1 s25 & 19.6 & -66.5 & 59.4 & $1.8\pm0.2$ & $1.6\pm0.5$ & $4.6\pm0.2$ & $2.8\pm0.2$ & $1.5\pm0.3$ & $1.0\pm0.2$ & -- & $10.12_{-0.09}^{+0.04}$ & $-0.05_{-0.06}^{+0.06}$ & $0.50_{-0.03}^{+0.00}$\\
inn1 s26 & 15.4 & -61.7 & 32.0 & $2.2\pm0.5$ & $1.4\pm1.0$ & $3.2\pm0.5$ & $2.9\pm0.5$ & $1.4\pm0.6$ & $1.5\pm0.5$ & -- & $10.12_{-0.24}^{+0.04}$ & $-0.37_{-0.23}^{+0.23}$ & $0.40_{-0.18}^{+0.08}$\\
inn1 s27 & 21.4 & -53.0 & 48.7 & $1.4\pm0.3$ & $5.1\pm0.6$ & $3.1\pm0.3$ & $2.7\pm0.3$ & $1.8\pm0.3$ & $2.6\pm0.3$ & -- & $10.15_{-0.16}^{+0.02}$ & $-0.16_{-0.10}^{+0.12}$ & $0.03_{-0.10}^{+0.10}$\\
inn1 s28 & 14.7 & -42.7 & 28.3 & $1.9\pm0.5$ & $1.4\pm1.1$ & $3.3\pm0.5$ & $1.2\pm0.5$ & $2.1\pm0.6$ & $1.5\pm0.4$ & -- & $10.14_{-0.20}^{+0.02}$ & $-0.52_{-0.23}^{+0.23}$ & $0.45_{-0.17}^{+0.05}$\\
inn1 s29 & 20.0 & -39.4 & 19.5 & $1.3\pm0.7$ & $4.0\pm2.0$ & $5.8\pm0.9$ & $1.5\pm0.8$ & $1.5\pm1.0$ & $0.6\pm0.8$ & -- & $10.15_{-0.21}^{+0.02}$ & $0.07_{-0.22}^{+0.22}$ & $0.49_{-0.12}^{+0.01}$\\
inn1 s30 & 14.4 & -22.7 & 30.2 & $0.7\pm0.5$ & $4.0\pm2.0$ & $3.8\pm0.7$ & $3.1\pm0.8$ & $5.5\pm0.9$ & $1.3\pm1.2$ & -- & $10.16_{-0.19}^{+0.02}$ & $0.29_{-0.22}^{+0.22}$ & $-0.18_{-0.09}^{+0.14}$\\
inn1 s31 & 19.6 & -24.8 & 21.2 & -- & $0.0\pm2.0$ & $5.2\pm0.8$ & $2.1\pm0.8$ & $2.5\pm1.1$ & $4.2\pm0.9$ & -- & $10.11_{-0.36}^{+0.05}$ & $-0.68_{-0.91}^{+0.91}$ & $0.16_{-0.25}^{+0.21}$\\
inn1 s32 & 14.1 & 2.5 & 43.1 & $1.7\pm0.3$ & $2.3\pm0.8$ & $5.2\pm0.3$ & $3.7\pm0.4$ & $2.9\pm0.4$ & $2.1\pm0.4$ & $0.7\pm0.2$ & $10.16_{-0.10}^{+0.01}$ & $0.27_{-0.10}^{+0.10}$ & $0.33_{-0.09}^{+0.09}$\\
inn1 s33 & 19.3 & -6.0 & 23.5 & $2.7\pm0.6$ & $1.0\pm2.0$ & $4.9\pm0.6$ & $5.2\pm0.7$ & $5.3\pm0.8$ & $0.7\pm0.7$ & -- & $10.02_{-0.25}^{+0.12}$ & $0.51_{-0.19}^{+0.19}$ & $0.06_{-0.16}^{+0.16}$\\
inn1 s34 & 16.4 & 31.7 & 26.6 & $1.0\pm0.6$ & $8.2\pm1.3$ & $4.0\pm0.5$ & $4.8\pm0.6$ & $4.8\pm0.7$ & $4.6\pm0.7$ & $1.4\pm0.4$ & $10.13_{-0.23}^{+0.03}$ & $0.73_{-0.13}^{+0.10}$ & $-0.27_{-0.03}^{+0.06}$\\
inn1 s35 & 19.0 & 21.0 & 36.2 & $1.7\pm0.4$ & $0.6\pm0.9$ & $3.1\pm0.3$ & $3.8\pm0.5$ & $5.1\pm0.6$ & $1.7\pm0.5$ & $1.4\pm0.3$ & $10.15_{-0.17}^{+0.02}$ & $0.01_{-0.14}^{+0.14}$ & $-0.23_{-0.05}^{+0.08}$\\
inn1 s36 & 21.0 & 41.4 & 24.8 & $1.8\pm0.5$ & $5.9\pm1.3$ & $2.5\pm0.5$ & $0.4\pm0.6$ & $1.8\pm0.8$ & $3.3\pm0.6$ & $1.4\pm0.4$ & $10.07_{-0.25}^{+0.09}$ & $-0.30_{-0.22}^{+0.22}$ & $0.00_{-0.17}^{+0.20}$\\
inn1 s37 & 23.7 & 35.8 & 20.8 & $3.6\pm0.8$ & -- & $5.3\pm0.6$ & -- & -- & $4.3\pm0.6$ & -- & $10.11_{-0.36}^{+0.05}$ & $-0.49_{-0.96}^{+0.86}$ & $0.10_{-0.23}^{+0.23}$\\
inn2 s18 & 21.7 & 53.6 & 22.3 & $2.2\pm0.7$ & $7.0\pm2.0$ & $10.8\pm0.8$ & $2.9\pm0.8$ & $1.0\pm0.9$ & $0.2\pm0.8$ & $-0.2\pm0.5$ & $10.13_{-0.11}^{+0.03}$ & $0.87_{-0.10}^{+0.03}$ & $0.50_{-0.03}^{+0.00}$\\
inn2 s19 & 21.2 & 65.0 & 17.9 & $0.4\pm0.8$ & $-2.0\pm2.0$ & $5.7\pm0.8$ & $4.3\pm0.8$ & $1.9\pm1.0$ & $1.5\pm0.9$ & $2.6\pm0.6$ & $10.11_{-0.36}^{+0.05}$ & $-0.66_{-0.90}^{+0.90}$ & $0.15_{-0.24}^{+0.22}$\\
inn2 s20 & 17.1 & 54.4 & 31.1 & -- & $5.0\pm2.0$ & -- & $2.5\pm0.8$ & $4.4\pm1.0$ & $1.1\pm1.1$ & $1.4\pm0.8$ & $10.11_{-0.36}^{+0.05}$ & $-0.66_{-0.88}^{+0.88}$ & $0.09_{-0.23}^{+0.23}$\\
inn2 s22 & 16.0 & 78.1 & 21.9 & $0.9\pm0.6$ & $5.4\pm1.4$ & $5.2\pm0.5$ & $2.2\pm0.8$ & $2.2\pm0.8$ & $0.9\pm0.6$ & $1.8\pm0.4$ & $10.14_{-0.19}^{+0.02}$ & $0.30_{-0.16}^{+0.16}$ & $0.46_{-0.11}^{+0.04}$\\
inn2 s23 & 21.2 & 98.8 & 13.1 & $1.8\pm0.9$ & $5.0\pm2.0$ & $4.9\pm1.0$ & $2.0\pm2.0$ & $2.0\pm2.0$ & $0.7\pm1.3$ & -- & $10.11_{-0.27}^{+0.05}$ & $0.16_{-0.30}^{+0.30}$ & $0.44_{-0.21}^{+0.06}$\\
inn2 s24 & 15.9 & 100.2 & 18.7 & $2.1\pm0.7$ & $3.0\pm2.0$ & $2.1\pm0.9$ & $-0.5\pm1.1$ & $3.4\pm1.2$ & $0.9\pm1.0$ & -- & $10.12_{-0.23}^{+0.04}$ & $-0.94_{-0.46}^{+0.46}$ & $0.25_{-0.25}^{+0.18}$\\
inn2 s25 & 20.7 & 112.2 & 15.2 & $0.1\pm0.8$ & $1.0\pm2.0$ & $3.9\pm0.9$ & -- & $-0.0\pm2.0$ & -- & -- & $10.11_{-0.35}^{+0.05}$ & $-0.69_{-0.90}^{+0.90}$ & $0.10_{-0.23}^{+0.23}$\\
inn2 s26 & 15.5 & 116.0 & 20.3 & $1.6\pm0.7$ & $-1.0\pm2.0$ & $2.8\pm0.8$ & $0.4\pm0.9$ & $1.8\pm1.3$ & $1.2\pm1.2$ & -- & $10.14_{-0.16}^{+0.02}$ & $-1.16_{-0.47}^{+0.47}$ & $0.42_{-0.26}^{+0.08}$\\
inn2 s28 & 14.3 & 138.7 & 27.9 & -- & $-0.9\pm1.1$ & $4.8\pm0.5$ & $2.4\pm0.6$ & $3.8\pm0.8$ & $1.2\pm0.6$ & -- & $10.11_{-0.35}^{+0.05}$ & $-0.39_{-0.97}^{+0.83}$ & $0.17_{-0.25}^{+0.21}$\\
inn2 s29 & 19.5 & 141.6 & 16.8 & $2.2\pm0.8$ & $4.0\pm2.0$ & $7.6\pm0.8$ & $3.2\pm0.9$ & $2.8\pm1.0$ & $-0.3\pm0.8$ & -- & $10.13_{-0.20}^{+0.03}$ & $0.58_{-0.20}^{+0.17}$ & $0.50_{-0.07}^{+0.00}$\\
inn2 s30 & 14.1 & 159.0 & 26.8 & $1.5\pm0.5$ & $5.4\pm1.3$ & $2.3\pm0.6$ & $1.6\pm0.7$ & $1.9\pm0.9$ & $1.2\pm0.8$ & -- & $10.14_{-0.21}^{+0.02}$ & $-0.47_{-0.24}^{+0.24}$ & $0.14_{-0.21}^{+0.20}$\\
inn2 s31 & 18.4 & 158.3 & 14.5 & $1.5\pm1.2$ & $3.0\pm3.0$ & $7.2\pm1.3$ & $-1.0\pm2.0$ & $3.0\pm2.0$ & -- & -- & $10.12_{-0.36}^{+0.04}$ & $-0.46_{-0.96}^{+0.84}$ & $0.11_{-0.23}^{+0.23}$\\
inn2 s32 & 13.6 & -175.1 & 25.3 & $1.2\pm0.7$ & $4.3\pm1.4$ & $4.4\pm0.7$ & $4.1\pm0.7$ & $4.0\pm0.8$ & $1.3\pm0.8$ & -- & $10.15_{-0.22}^{+0.02}$ & $0.33_{-0.20}^{+0.21}$ & $0.03_{-0.16}^{+0.18}$\\
inn2 s33 & 18.0 & 178.0 & 15.1 & $-0.1\pm0.9$ & $4.0\pm2.0$ & $1.5\pm1.0$ & $0.9\pm1.0$ & $1.3\pm1.2$ & $-0.3\pm1.0$ & -- & $10.15_{-0.14}^{+0.02}$ & $-1.01_{-0.41}^{+0.41}$ & $0.24_{-0.26}^{+0.19}$\\
inn2 s35 & 19.0 & -162.9 & 16.5 & $1.9\pm0.8$ & $4.0\pm2.0$ & $2.4\pm0.8$ & $4.2\pm0.9$ & $-0.8\pm1.1$ & $-0.1\pm1.0$ & -- & $10.13_{-0.28}^{+0.04}$ & $-0.49_{-0.36}^{+0.36}$ & $0.31_{-0.25}^{+0.15}$\\
inn2 s36 & 19.8 & -143.5 & 19.7 & $1\pm34$ & $2.0\pm2.0$ & $3.6\pm0.8$ & $-0.7\pm0.9$ & $2.4\pm1.0$ & $3.8\pm0.8$ & $1.3\pm0.5$ & $10.14_{-0.32}^{+0.03}$ & $-0.34_{-0.30}^{+0.31}$ & $0.38_{-0.23}^{+0.11}$\\
out1 s13 & 34.6 & -131.6 & 9.9 & $1.6\pm0.9$ & -- & $8.3\pm1.3$ & $4.6\pm1.4$ & -- & $5.0\pm1.2$ & $-5.6\pm0.8$ & $10.11_{-0.35}^{+0.05}$ & $-0.66_{-0.90}^{+0.90}$ & $0.10_{-0.23}^{+0.23}$\\
out1 s15 & 36.2 & -105.9 & 17.0 & $2.4\pm1.1$ & $7.0\pm2.0$ & $4.9\pm1.0$ & $4.0\pm1.2$ & $4.0\pm2.0$ & $5.0\pm2.0$ & $2.2\pm0.9$ & $10.07_{-0.25}^{+0.08}$ & $0.68_{-0.24}^{+0.16}$ & $0.01_{-0.19}^{+0.21}$\\
out1 s16 & 30.2 & -113.4 & 13.5 & $1.2\pm1.2$ & $1.0\pm3.0$ & $6.8\pm1.2$ & $0.2\pm1.4$ & $7.0\pm2.0$ & $3.6\pm1.5$ & $1.1\pm0.8$ & $10.15_{-0.21}^{+0.02}$ & $0.52_{-0.31}^{+0.24}$ & $0.40_{-0.22}^{+0.09}$\\
out1 s17 & 36.6 & -92.7 & 17.3 & $2.0\pm0.8$ & $-4.0\pm2.0$ & $3.8\pm0.8$ & $6.3\pm0.8$ & $4.0\pm0.9$ & $2.1\pm0.7$ & -- & $10.16_{-0.22}^{+0.02}$ & $0.09_{-0.25}^{+0.25}$ & $-0.09_{-0.14}^{+0.19}$\\
out1 s18 & 30.5 & -96.3 & 40.8 & $2.4\pm0.4$ & $5.6\pm0.7$ & $2.4\pm0.3$ & $1.4\pm0.4$ & $2.1\pm0.4$ & $1.8\pm0.3$ & -- & $9.74_{-0.24}^{+0.22}$ & $-0.24_{-0.17}^{+0.17}$ & $0.07_{-0.14}^{+0.14}$\\
out1 s19 & 36.8 & -82.8 & 43.6 & $2.7\pm0.3$ & $4.9\pm0.7$ & $3.7\pm0.3$ & $2.4\pm0.3$ & $2.3\pm0.4$ & $0.9\pm0.3$ & -- & $9.40_{-0.16}^{+0.20}$ & $0.14_{-0.15}^{+0.18}$ & $0.45_{-0.08}^{+0.04}$\\
out1 s20 & 30.5 & -83.8 & 82.2 & $1.6\pm0.2$ & $4.6\pm0.4$ & $4.4\pm0.2$ & $3.1\pm0.2$ & $2.5\pm0.2$ & $1.33\pm0.14$ & -- & $10.15_{-0.10}^{+0.02}$ & $0.05_{-0.06}^{+0.08}$ & $0.30_{-0.05}^{+0.05}$\\
out1 s21 & 37.2 & -73.4 & 47.0 & $1.6\pm0.2$ & $5.1\pm0.6$ & $4.3\pm0.3$ & $3.3\pm0.3$ & $1.9\pm0.4$ & $1.6\pm0.2$ & -- & $10.09_{-0.17}^{+0.07}$ & $0.16_{-0.12}^{+0.12}$ & $0.31_{-0.08}^{+0.08}$\\
out1 s22 & 37.3 & -68.9 & 36.4 & $0.6\pm0.4$ & $3.6\pm0.9$ & $5.2\pm0.4$ & $3.9\pm0.4$ & $1.9\pm0.5$ & $1.9\pm0.4$ & -- & $10.16_{-0.09}^{+0.01}$ & $0.31_{-0.11}^{+0.11}$ & $0.38_{-0.10}^{+0.08}$\\
out1 s23 & 32.8 & -60.6 & 48.8 & $0.8\pm0.3$ & $5.5\pm0.5$ & $4.7\pm0.3$ & $2.6\pm0.3$ & $2.2\pm0.3$ & $2.2\pm0.3$ & -- & $10.16_{-0.07}^{+0.02}$ & $0.13_{-0.08}^{+0.09}$ & $0.31_{-0.08}^{+0.08}$\\
out1 s24 & 35.6 & -53.4 & 20.1 & $1.5\pm0.6$ & $6.0\pm2.0$ & $7.8\pm0.7$ & $2.7\pm0.8$ & $3.6\pm0.9$ & $-0.0\pm0.7$ & -- & $10.13_{-0.18}^{+0.03}$ & $0.74_{-0.17}^{+0.11}$ & $0.49_{-0.07}^{+0.01}$\\
out1 s25 & 27.6 & -45.7 & 28.5 & $1.7\pm0.4$ & $4.1\pm1.0$ & $4.1\pm0.5$ & $4.4\pm0.5$ & $2.4\pm0.6$ & $1.6\pm0.5$ & -- & $10.11_{-0.24}^{+0.05}$ & $0.19_{-0.16}^{+0.19}$ & $0.16_{-0.14}^{+0.14}$\\
out1 s27 & 34.5 & -38.5 & 10.5 & $2.0\pm2.0$ & $-6.0\pm4.0$ & $0.0\pm3.0$ & $1.0\pm4.0$ & $0.0\pm5.0$ & $-1.0\pm7.0$ & -- & $10.11_{-0.28}^{+0.05}$ & $-1.73_{-0.41}^{+0.63}$ & $0.21_{-0.25}^{+0.20}$\\
out1 s28 & 28.0 & -28.9 & 10.0 & $-1.1\pm1.4$ & $5.0\pm4.0$ & $8.0\pm2.0$ & $2.0\pm3.0$ & $7.0\pm3.0$ & $1.0\pm4.0$ & -- & $10.15_{-0.21}^{+0.02}$ & $0.86_{-0.38}^{+0.04}$ & $0.24_{-0.25}^{+0.18}$\\
out1 s29 & 34.0 & -28.5 & 8.0 & $4.5\pm1.5$ & $5.0\pm4.0$ & $4.0\pm3.0$ & $1.0\pm3.0$ & $2.0\pm4.0$ & $1.0\pm4.0$ & -- & $9.91_{-0.37}^{+0.21}$ & $-0.74_{-0.79}^{+0.79}$ & $0.26_{-0.26}^{+0.18}$\\
out1 s30 & 26.7 & -14.0 & 13.7 & $1.3\pm1.1$ & $-1.0\pm3.0$ & $4.6\pm1.4$ & $1.9\pm1.5$ & $5.0\pm2.0$ & $4.0\pm2.0$ & -- & $10.15_{-0.26}^{+0.02}$ & $0.08_{-0.44}^{+0.40}$ & $0.22_{-0.24}^{+0.19}$\\
out1 s31 & 33.3 & -17.5 & 6.7 & $-2.0\pm2.0$ & $12.0\pm4.0$ & $-3.0\pm2.0$ & $-2.0\pm2.0$ & $1.0\pm3.0$ & $-1.0\pm3.0$ & -- & $10.15_{-0.22}^{+0.02}$ & $-1.05_{-0.60}^{+0.64}$ & $0.02_{-0.21}^{+0.24}$\\
out1 s32 & 26.6 & 3.2 & 23.0 & $1.6\pm0.5$ & $2.6\pm1.3$ & $4.7\pm0.5$ & $2.3\pm0.7$ & $1.9\pm0.7$ & $3.3\pm0.6$ & -- & $10.14_{-0.19}^{+0.02}$ & $0.12_{-0.16}^{+0.18}$ & $0.41_{-0.14}^{+0.07}$\\
out1 s34 & 28.0 & 27.7 & 23.0 & $1.3\pm1.4$ & $3.0\pm2.0$ & $2.4\pm1.5$ & $-0.0\pm2.0$ & $1.0\pm3.0$ & $4.0\pm3.0$ & $3.0\pm2.0$ & $10.11_{-0.35}^{+0.05}$ & $-0.70_{-0.90}^{+0.90}$ & $0.11_{-0.23}^{+0.23}$\\
out1 s35 & 33.0 & 15.1 & 9.8 & $3.0\pm2.0$ & $4.0\pm3.0$ & $-0.3\pm1.4$ & $1.0\pm2.0$ & $4.0\pm2.0$ & -- & $2.1\pm0.9$ & $10.11_{-0.36}^{+0.05}$ & $-0.41_{-0.98}^{+0.84}$ & $0.10_{-0.23}^{+0.23}$\\
out1 s36 & 33.3 & 37.1 & 15.1 & $1.5\pm0.2$ & $6.0\pm3.0$ & $2.9\pm1.2$ & $3.0\pm2.0$ & -- & $1.0\pm2.0$ & $0.1\pm1.0$ & $10.11_{-0.36}^{+0.05}$ & $-0.68_{-0.91}^{+0.91}$ & $0.10_{-0.23}^{+0.23}$\\
out2 s14 & 37.0 & 71.6 & 12.5 & $2.0\pm2.0$ & $9.0\pm4.0$ & $3.0\pm2.0$ & $3.0\pm3.0$ & $-3.0\pm4.0$ & $0.0\pm4.0$ & $0.0\pm3.0$ & $10.09_{-0.34}^{+0.07}$ & $0.20_{-0.69}^{+0.48}$ & $0.14_{-0.24}^{+0.22}$\\
out2 s15 & 29.3 & 56.7 & 22.6 & $1.1\pm0.6$ & $3.0\pm2.0$ & $3.6\pm0.7$ & $3.3\pm0.7$ & $1.1\pm0.9$ & $1.6\pm0.8$ & $1.2\pm0.5$ & $10.15_{-0.23}^{+0.02}$ & $-0.07_{-0.24}^{+0.24}$ & $0.33_{-0.21}^{+0.13}$\\
out2 s16 & 37.1 & 84.2 & 12.2 & $1.7\pm1.2$ & $11.0\pm2.0$ & $0.4\pm1.2$ & $1.8\pm1.5$ & $1.0\pm2.0$ & $4.1\pm1.5$ & -- & $10.01_{-0.37}^{+0.14}$ & $-0.14_{-0.44}^{+0.44}$ & $-0.19_{-0.09}^{+0.18}$\\
out2 s17 & 30.4 & 79.2 & 14.8 & $2.3\pm0.7$ & $4.0\pm2.0$ & $2.7\pm0.9$ & $1.0\pm0.9$ & $0.2\pm1.2$ & $1.4\pm1.0$ & $1.6\pm0.7$ & $10.08_{-0.28}^{+0.08}$ & $-0.57_{-0.38}^{+0.38}$ & $0.40_{-0.25}^{+0.09}$\\
out2 s18 & 37.0 & 94.3 & 12.5 & $1.5\pm0.9$ & $-0.0\pm2.0$ & $1.9\pm1.2$ & $4.0\pm1.0$ & $1.9\pm1.2$ & -- & -- & $10.12_{-0.36}^{+0.04}$ & $-0.68_{-0.89}^{+0.89}$ & $0.17_{-0.25}^{+0.21}$\\
out2 s19 & 29.8 & 92.5 & 16.3 & $2.4\pm1.0$ & $5.0\pm3.0$ & $1.4\pm1.4$ & $-0.0\pm2.0$ & $2.0\pm2.0$ & $-1.0\pm2.0$ & $1.0\pm2.0$ & $10.10_{-0.30}^{+0.07}$ & $-1.01_{-0.59}^{+0.62}$ & $0.19_{-0.25}^{+0.20}$\\
out2 s20 & 34.9 & 100.9 & 14.9 & $2.5\pm1.0$ & $10.0\pm2.0$ & $1.3\pm1.2$ & $4.4\pm1.4$ & $4.0\pm2.0$ & -- & -- & $10.11_{-0.37}^{+0.05}$ & $-0.43_{-0.96}^{+0.83}$ & $0.18_{-0.25}^{+0.21}$\\
out2 s21 & 29.5 & 101.5 & 16.0 & $1.6\pm0.9$ & $5.0\pm2.0$ & $1.3\pm1.1$ & $4.5\pm1.2$ & $2.0\pm2.0$ & $2.2\pm1.2$ & -- & $10.11_{-0.32}^{+0.06}$ & $-0.21_{-0.40}^{+0.43}$ & $-0.20_{-0.09}^{+0.18}$\\
out2 s22 & 36.6 & 110.6 & 6.3 & $4.0\pm2.0$ & $5.0\pm6.0$ & $-2.0\pm4.0$ & -- & $2.0\pm7.0$ & -- & -- & $10.11_{-0.36}^{+0.05}$ & $-0.70_{-0.91}^{+0.91}$ & $0.19_{-0.25}^{+0.21}$\\
out2 s26 & 27.6 & 133.4 & 14.4 & $1.3\pm1.0$ & $-0.0\pm3.0$ & $6.6\pm1.3$ & $5.0\pm2.0$ & $6.0\pm2.0$ & $1.0\pm2.0$ & -- & $10.14_{-0.22}^{+0.02}$ & $0.67_{-0.31}^{+0.18}$ & $0.37_{-0.23}^{+0.12}$\\
out2 s28 & 27.2 & 144.8 & 12.5 & $1.8\pm1.1$ & $-2.0\pm3.0$ & $5.3\pm1.4$ & $2.0\pm2.0$ & $0.0\pm2.0$ & $3.0\pm2.0$ & -- & $10.14_{-0.26}^{+0.02}$ & $-0.20_{-0.64}^{+0.55}$ & $0.45_{-0.26}^{+0.05}$\\
out2 s29 & 33.3 & 146.7 & 7.8 & $5.6\pm1.4$ & -- & $-4.0\pm2.0$ & $5.0\pm2.0$ & $-0.0\pm2.0$ & $0.2\pm1.5$ & -- & $10.11_{-0.35}^{+0.05}$ & $-0.39_{-0.99}^{+0.83}$ & $0.10_{-0.23}^{+0.23}$\\
out2 s30 & 26.6 & 159.1 & 17.8 & $1.9\pm1.0$ & $3.0\pm2.0$ & $2.5\pm1.4$ & $1.0\pm2.0$ & $-0.0\pm2.0$ & $-1.0\pm2.0$ & -- & $10.13_{-0.25}^{+0.03}$ & $-1.03_{-0.58}^{+0.62}$ & $0.29_{-0.27}^{+0.16}$\\
out2 s31 & 33.2 & 157.0 & 10.0 & $0.3\pm1.3$ & $12.0\pm4.0$ & $-4.0\pm2.0$ & $2.0\pm2.0$ & $-2.0\pm3.0$ & $1.0\pm2.0$ & -- & $10.13_{-0.23}^{+0.03}$ & $-1.40_{-0.48}^{+0.57}$ & $-0.06_{-0.18}^{+0.24}$\\
out2 s32 & 26.2 & 174.8 & 17.6 & $1.8\pm1.0$ & $7.0\pm3.0$ & $5.0\pm2.0$ & $3.0\pm2.0$ & $6.0\pm3.0$ & $0.0\pm3.0$ & -- & $10.11_{-0.28}^{+0.05}$ & $0.63_{-0.43}^{+0.23}$ & $0.10_{-0.22}^{+0.22}$\\
out2 s33 & 32.6 & 169.5 & 17.0 & $5.3\pm0.9$ & -- & $2.6\pm0.9$ & $5.0\pm0.9$ & $2.0\pm1.0$ & $2.5\pm0.9$ & -- & $10.11_{-0.36}^{+0.05}$ & $-0.48_{-0.96}^{+0.85}$ & $0.10_{-0.23}^{+0.23}$\\
out2 s34 & 25.9 & -159.4 & 23.5 & $3.0\pm2.0$ & $1.0\pm2.0$ & $-0.3\pm0.9$ & $4.3\pm0.8$ & $-2.6\pm1.3$ & $-0.5\pm1.1$ & -- & $10.03_{-0.32}^{+0.12}$ & $-1.24_{-0.39}^{+0.39}$ & $-0.06_{-0.17}^{+0.23}$\\
out2 s35 & 27.2 & -145.8 & 21.1 & -- & $2.1\pm0.6$ & $1.0\pm0.7$ & $7.9\pm0.8$ & $3.6\pm1.0$ & $4.0\pm0.8$ & -- & $10.16_{-0.29}^{+0.02}$ & $-0.57_{-0.19}^{+0.19}$ & $-0.27_{-0.03}^{+0.08}$\\